\documentclass[12pt,a4paper]{article}
\usepackage{a4,epsf,here}
\def\K{\hbox{\bf K}}

\def\sn{{\rm sn}}
\def\dn{{\rm dn}}
\def\cn{{\rm cn}}
\def\am{{\rm am}}
\def\tx{\dot x}

\def\TA{T_{\rm A}}
\def\MA{{\rm M}_{\rm A}}
\def\trMA{{\rm tr}\,\MA}
\def\trM{{\rm tr\,M}}

\def\be{\begin{equation}}
\def\ee{\end{equation}}
\def\bea{\begin{eqnarray}}
\def\eea{\end{eqnarray}}
\def\eq#1{(\ref{#1})}
\def\eqq#1#2{(\ref{#1},\ref{#2})}
\def\bs{\bigskip}
\def\ms{\medskip}
\def\fig#1{Fig.~\ref{#1}}
\def\tab#1{Table \ref{#1}}
\def\etal{{\it et al}}
\newcommand{\Figurebb}[9]{
\begin{figure}[H]
\leavevmode
\epsfysize=#7cm
\epsfbox[#2 #3 #4 #5]{#6}
\par
\parbox{#8cm}{
\caption[figure]{\renewcommand{\baselinestretch}{0.8} \small
                                           \hspace{-0.3truecm}#9}
\label{#1}}
\end{figure}
}
\newcommand{\Table}[4]{
\begin{table}[H]\begin{center}{#3}
\parbox{#2cm}{
\vspace{0.5cm}
\caption[table]{\renewcommand{\baselinestretch}{0.8} \small
                                           \hspace{-0.3truecm}#4}
\label{#1}}
\end{center}
\end{table}
}

\begin{document}

\baselineskip 14pt

\centerline{\bf \Large Occurrence of periodic Lam\'e functions}

\bs

\centerline{\bf \Large at bifurcations in chaotic Hamiltonian systems}

\bs
\bs
\bs

\centerline{\bf M Brack$^1$, M Mehta$^1$ and K Tanaka$^{1,2}$}

\ms
\bs

\centerline{\it $^1$Institute for Theoretical Physics, University of
Regensburg, D-93040 Regensburg, Germany}

\centerline{\it $^2$Dept.\ of Physics, University of Saskatchewan, 
Saskatoon, SK, Canada S7N5E2}

\bs
\bs
\bs

\centerline{\bf \large Abstract}

\bs

\baselineskip 10pt
\small{
\noindent
We investigate cascades of isochronous pitchfork bifurcations of 
straight-line librating orbits in some two-dimensional Hamiltonian systems 
with mixed phase space. We show that the new bifurcated orbits, which are 
responsible for the onset of chaos, are given analytically by the periodic 
solutions of the Lam\'e equation as classified in 1940 by Ince. In 
Hamiltonians with C$_{2v}$ symmetry, they occur alternatingly as Lam\'e 
functions of period 2{\bf K} and 4{\bf K}, respectively, where 4{\bf K} is 
the period of the Jacobi elliptic function appearing in the Lam\'e equation. 
We also show that the two pairs of orbits created at period-doubling 
bifurcations of island-chain type are given by two different linear 
combinations of algebraic Lam\'e functions with period 8{\bf K}.
}

\baselineskip 14pt

\vspace*{1cm}

\section{Introduction}

\noindent
One of the well-established routes to chaos in maps is the
so-called Feigenbaum scenario \cite{feig} which consists in a cascade
of successive period-doubling bifurcations of pitchfork type. They
were first discussed for the 1-dimensional logistic map by Feigenbaum 
\cite{feig} and then also found in the area-conserving two-dimensional
H\'enon map \cite{fei2,fei3}, although the numerical scaling constants 
found there differ from those in the 1-dimensional case. One of us (M.B.)
has recently investigated \cite{mbgu} similar cascades of pitchfork
bifurcations occurring in 2-dimensional Hamiltonian systems with mixed 
dynamics, whereby the scaling constants can be determined analytically 
and depend on the potential parameters. In the present paper, we shall
show that the new orbits born at its bifurcations are, near the 
bifurcations points, analytically given by periodic solutions of a 
linear second-order differential equation studied over 160 years ago by 
G. Lam\'e \cite{lame}, and therefore called the ``periodic Lam\'e 
functions'' \cite{strt}. They have been classified uniquely in 1940 by 
Ince \cite{inc1} who also derived their Fourier series expansions 
\cite{inc2}. We find that these not only reproduce accurately the periodic 
orbits found numerically by solving the equations of motion at the
bifurcations, but in the H\'enon-Heiles \cite{hh} and similar potentials 
the Lam\'e functions can also be used to describe the evolution of the 
bifurcated orbits at higher energies. A particularly interesting case is 
the homogeneous quartic oscillator for which the Lam\'e functions become
finite polynomials in terms of Jacobi elliptic functions. Here we can also 
find analytical expressions for the algebraic Lam\'e functions which 
describe the orbits created at period-doubling bifurcations of island-chain 
type.

\vfill
\newpage

\section{Bifurcations of a straight-line librating orbit}
\label{stasec}

\noindent
We start from an autonomous two-dimensional Hamiltonian of a particle
with unit mass in a smooth potential $V(x,y)$
\be
H = \frac12 \left( p_x^2 + p_y^2 \right) + V(x,y)\,.
\ee
Assume that there exists a straight-line librating orbit, called A, along 
the $y$ axis, so that
\be
x_A(t)\equiv0\,, \qquad y_A(t)=y_A(t+\TA)
\ee
are solutions of the equations of motion and $\TA$ is the period of the A
orbit. Its stability is obtained from the stability matrix $\MA$ that 
describes the propagation of the linearized flow of a small perturbation 
$\delta x(t)$, $\delta p_x(t)=\delta\tx(t)$ transverse to the orbit A:
\be
\left(\begin{array}{r} \delta   x\,(\TA)\\
                       \delta p_x\,(\TA)
      \end{array} \right)
   = \MA \left(\begin{array}{r} \delta   x\,(0)\\
                                \delta p_x\,(0)
      \end{array} \right).
\ee
When $-2 < \trMA < +2$, the orbit is stable, for $|\trMA|>2$ it is unstable. 
Marginally stable orbits with $\trMA=+2$ occur in systems with continuous 
symmetries; in two dimensions this would imply integrability. We 
investigate here only non-integrable systems in which all orbits are 
isolated. Then, an orbit must undergo a bifurcation when $\trMA=+2$. 

The elements of the stability matrix $\MA$ can be calculated from solutions 
of the linearized equation of motion in the transverse $x$ direction, which 
we write in the Newtonian form
\be 
\left.
\delta {\ddot x}(t) + \frac{\partial^{\,2} V(x,y)}{\partial x^{2}}
               \right|_{x=0,y=y_A(t)} \delta x(t) = 0\,.
\ee 
This equation is identical to Hill's equation \cite{hill} in its standard 
form \cite{mawi}
\be
\delta {\ddot x}(t) + [\lambda + Q(t)]\, \delta x(t) = 0\,,
\label{hill}
\ee
where $Q(t)$ is a $\TA$ (or $\TA/2$) periodic function whose constant 
Fourier component is zero. In general, the solutions of \eq{hill} are 
non-periodic. However, periodic solutions with period $\TA$ (or $\TA/2$) 
and multiples thereof exist for specific discrete values of $\lambda$. This 
happens exactly at bifurcations of the A orbit where $\trMA=+2$. The 
periodic solutions $\delta x(t)$ found at these discrete values of 
$\lambda$ describe the $x$ motion of the bifurcated orbits infinitely close 
to the bifurcation point. A special case of the Hill equation is the Lam\'e 
equation which we discuss in the following section.

\section{The periodic Lam\'e functions}
\label{lamsec}

\noindent
One of the standard forms of the Lam\'e equation reads \cite{erde}
\be
\Lambda''(z) + \left[h - n(n+1)\,k^2\,{\rm \sn}^2(z,k)\right] \Lambda(z) 
             = 0\,,
\label{lameq}
\ee
where $\sn(z,k)$ is a Jacobi elliptic function with modulus $k$ limited 
by $0\le k < 1$. The real period of $\sn(z,k)$ in the variable $z$ is 
4{\bf K}, where
\be
\K = K(k) = F\left(\frac{\pi}{2},k\right)
\ee
is the complete elliptic integral of the first kind with modulus $k$. 
We follow throughout this paper the notation of Gradshteyn and Ryzhik 
\cite{gr} for elliptic functions and integrals. We are interested here 
only in real solutions for $\Lambda(z)$ with real argument $z$. 
Hence $h$ and $n(n+1)$ are assumed here to be arbitrary real constants. 
This means that $n$ is either real, or complex with real part $-\frac12$. 
There is a vast literature on the periodic solutions of Eq.\ \eq{lameq}; 
for an exhaustive presentation of their definition and series expansions 
as well as the most relevant literature, we refer to Erd\'elyi {\it 
et al}\ \cite{erde}. (See also \cite{strt} for literature prior to 1932.) 
Ince \cite{inc1,inc2} has introduced a unique classification and 
nomenclature for the four types of periodic solutions, calling them 
Ec$_n^m(z)$ and Ec$_n^m(z)$, where $n$ is the parameter appearing in 
\eq{lameq} and $m$ an integer giving the number of zeros in the interval 
$0\leq z < 2\K$. Following a slight redefinition by Erd\'elyi \cite{erd1}, 
the Ec$(z)$ are even and the Es$(z)$ are odd functions of $z-\K$, 
respectively. When $m$ is an even integer, the Lam\'e functions have the 
period $2\K$ in the variable $z$, which is the same as the period of 
$\sn^2(z,k)$ appearing in \eq{lameq}; when $m$ is odd, they have the 
period $4\K$. Solutions with period $2p\K$ ($p=3,4,\dots$) can also be 
found; we shall discuss some solutions with period $8\K$ further below. 
All these periodic solutions exist only for discrete eigenvalues of $h$, 
denoted by $a_n^m$ and $b_n^m$ for the Ec$_n^m$ and Es$_n^m$, respectively; 
there exists exactly one solution of each of the above four types of Lam\'e 
functions for each $m\geq 0$. The eigenvalues of $h$ can, in principle, be 
found by solving the characteristic equation obtained from an infinite 
continued fraction \cite{inc1} which is, however, rather difficult to 
evaluate in 
the general case. In the context of our paper, they are determined by 
bifurcations of a linear periodic orbit and we obtain them therefore from 
a numerical calculation of its stability discriminant $\trMA$.

The Fourier expansions derived by Ince \cite{inc2}, with the modification 
by Erd\'elyi \cite{erd1}, are given in terms of the variable
\be
\zeta = \frac{\pi}{2} - \am(z,k),
\label{zeta}
\ee
where $\am(z,k)=\arcsin[\sn(z,k)]$, and read as follows:
\begin{eqnarray}
{\rm Ec}_n^{2m}(z)  &=&\frac12 A_0 + \sum_{r=1}^\infty A_{2r}\cos(2r\zeta)
                       \,,\qquad\qquad\qquad(\hbox{period }2\K)\label{ece}\\
{\rm Ec}_n^{2m+1}(z)&=&\sum_{r=0}^\infty A_{2r+1}\cos[(2r+1)\zeta]\,,\!
                       \qquad\qquad\qquad (\hbox{period }4\K) \label{eco}\\
{\rm Es}_n^{2m}(z)  &=&\sum_{r=1}^\infty B_{2r}\sin(2r\zeta)\,,\qquad\quad\;
                       \qquad\qquad\qquad (\hbox{period }2\K)\label{ese}\\
{\rm Es}_n^{2m+1}(z)&=&\sum_{r=0}^\infty B_{2r+1}\sin[(2r+1)\zeta]\,,
                       \qquad\qquad\qquad (\hbox{period } 4\K) \label{eso}
\end{eqnarray}
with $m=0,$ 1, 2, \dots The expansion coefficients can be calculated by 
two-step recurrence relations; we give them here only for the $A_{2r}$ 
\bea
\left[n(n+1)-2\right]k^2 A_2 & = & [2h-k^2n(n+1)] A_0\,, \label{recrel0}\\
\left[n(n+1)-(2r+2)(2r+1)\right]k^2A_{2r+2} & = & 
            2\!\left[2h-k^2n(n+1)-4r^2(2-k^2)\right] A_{2r} \nonumber\\ 
       & & - \left[n(n+1)-(2r-2)(2r-1)\right]k^2 A_{2r-2}\,,\qquad
\label{recrel}
\eea
(with $r=1,$ 2, 3, \dots) and refer to Erd\'elyi \etal\ (\cite{erde}, 
ch 15.5.1) for the other recurrence relations which look very similar. 

Although the series \eq{ece} - \eq{eso} are known \cite{inc2} to converge 
for $k<1$, they turned out to be semiconvergent in our numerical 
calculations for the cases with complex $n$, due to the fact that the 
characteristic values of $h$ were only determined approximately. We have 
truncated the above series at the value $r_{max}$ where the corresponding 
coefficient has its smallest absolute value before starting to diverge. 
The cut-off values $r_{max}$ were found to increase with the order $m$ 
of the Lam\'e function; their values are given in the Tables \ref{hhlame} 
and \ref{r4lame} in Secs.\ \ref{hhsec} and \ref{h4sec}, respectively.

When $n$ is an integer, the Fourier series terminate at finite values of 
$r$. The Lam\'e functions then become \cite{inc1} finite polynomials in 
the Jacobi elliptic functions $\sn(z)$, $\dn(z)$, and $\cn(z)$, and are 
called the ``Lam\'e polynomials'' in short. In Sec.\ \ref{q4sec} we will 
encounter a special case of the Lam\'e equation in which $h$ and $n$ are 
not independent, but where $h=\frac12 n(n+1)$ along with $k^2=\frac12$. 
Then, to each integer $n$ there exists only one value of $m$. Although 
the lowest few polynomials of this type and their eigenvalues of $h$ are 
included in the tables given by Ince \cite{inc1}, we give below their 
explicit expressions which take a particularly simple form. The basic four 
types of solutions correspond to the four rest classes modulo 4 of the 
integer $n$. With $p=0,$ 1, 2, 3, we obtain the following sums which are 
finite since the expansion coefficients become identically zero for $r>p$:
\bea
{\rm Ec}_{4p}^{2p}(z)&=&\sum_{r=0}^p A_{4r}\,\cn^{4r}(z)\,,\qquad\qquad
                        \qquad\qquad\qquad (\hbox{period } 2\K)\label{lce}\\
{\rm Ec}_{4p+1}^{2p+1}(z)&=&\cn(z)\sum_{r=0}^p C_{4r}\,\cn^{4r}(z)\,,\quad\;\;
                            \qquad\qquad\qquad(\hbox{period }4\K)\label{lco}\\
{\rm Es}_{4p+2}^{2p+1}(z)&=&\dn(z)\,\sn(z) \sum_{r=0}^p D_{4r}\,\cn^{4r}(z)\,,
                            \qquad\qquad\quad(\hbox{period }4\K)\label{lso}\\
{\rm Es}_{4p+3}^{2p+2}(z)&=&\cn(z)\,\dn(z)\,\sn(z)\sum_{r=0}^p B_{4r}\,
                            \cn^{4r}(z)\,.
                            \quad\quad\;\;\,(\hbox{period }2\K)\label{lse}
\eea
The simple one-step recurrence relations for the coefficients are (with 
$r=0,$ 1, 2, \dots, $p-1$)
\bea
(r+1)(4r+3)\,A_{4r+4} & = & - [p(4p+1)-r(4r+1)]\, A_{4r}\,,\\
2(r+1)(4r+5)\,C_{4r+4} & = & - [(2p+1)(4p+1)-(2r+1)(4r+1)]\, C_{4r}\,,\\
2(r+1)(4r+3)\,D_{4r+4} & = & - [(2p+1)(4p+3)-(2r+1)(4r+3)]\, D_{4r}\,,\\
(r+1)(4r+5)\,B_{4r+4} & = & - [(p+1)(4p+3)-(r+1)(4r+3)]\, B_{4r}\,.
\eea 
The first sixteen Lam\'e polynomials obtained from the above equations 
are given explicitly in Sec.\ \ref{q4sec} (see \tab{q4lame}), with the 
normalization $A_0=B_0=C_0=D_0=1$.

For half-integer values of $n$, one obtains periodic Lam\'e functions 
with period $8\K$ that have algebraic forms in the Jacobi elliptic 
integrals and are called ``algebraic Lam\'e functions'' \cite{inc2,erd2}. 
For the special case with $k^2=\frac12$ and $h=\frac12 n(n+1)$ we will  
encounter them in Sec.\ \ref{q4sec} at period-doubling bifurcations of 
island-chain type. As shown in a beautiful paper by Ince \cite{inc2}, 
there exist two linearly independent periodic solutions for 
$n=2p+\frac12$ with $p=0,$ 1, 2, \dots 
\bea
{\rm Ec}_{2p+1/2}^{m+1/2}(z) & = & \sqrt{\dn(z)+\cn(z)}\,
         \left\{\sum_{r=0}^p A_r\,\sn^{2r}(z)+\cn(z)\,\dn(z)
                \sum_{r=0}^{p-1}B_r\,\sn^{2r}(z)\right\},\label{ecalg}\\
{\rm Es}_{2p+1/2}^{m+1/2}(z) & = & \sqrt{\dn(z)-\cn(z)}\,
         \left\{\sum_{r=0}^p A_r\,\sn^{2r}(z)-\cn(z)\,\dn(z)
                \sum_{r=0}^{p-1}B_r\,\sn^{2r}(z)\right\},\label{esalg}
\eea  
where $m$ is the number of zeros in the open interval $(0,2\K)$; the 
coefficients $A_r$ and $B_r$ are given by two coupled recurrence relations. 
For $k^2=\frac12$, $h=\frac12 n(n+1)=2p(p+1)+\frac38$, there is only one
solution with $m=p$ for each $p$, and the recurrence relations read
\bea
2(2r+2)^2A_{r+1}+[4p(p+1)-6r(2r+1)]A_r-4(p-r+1)(p+r)A_{r-1}\qquad\;\;
                                                              \nonumber\\
  = 2(2r+2)B_{r+1}-3(2r+1)B_r+2rB_{r-1},\hspace*{-1cm}\label{alcof1}\\
(2r+2)^2B_{r+1}+[2p(p+1)-3(r+1)(2r+1)]B_r-2(p-r)(p+r+1)B_{r-1}\nonumber\\
  = (2r+2)A_{r+1}.\hspace*{-1cm}\label{alcof2}
\eea
These relations hold for $r\geq 0$ provided that coefficients with negative
indices $r$ are taken to be zero. It is quite easy to see that 
$B_r=A_{r+1}=0$ for $r\geq p$, which justifies the upper limits of the sums 
above. The coefficient $A_0$ may be used for the overall normalization of
both functions.

The algebraic Lam\'e functions \eq{ecalg} and \eq{esalg} are even and odd 
functions of $z$, respectively, and related to each other by 
Ec$_{2p+1/2}^{m+1/2}(z+2\K)=$ Es$_{2p+1/2}^{m+1/2}(z)$, which amounts to a 
sign change in front of $\cn(z)$. Erd\'elyi \cite{erd2} showed that the 
linear combinations Ec$_{2p+1/2}^{m+1/2}(z)\;+$ Es$_{2p+1/2}^{m+1/2}(z)$ 
and Ec$_{2p+1/2}^{m+1/2}(z)\;-$ Es$_{2p+1/2}^{m+1/2}(z)$ are even and odd
functions of $z-\K$, respectively, and proposed that they be used instead of 
the functions \eqq{ecalg}{esalg} introduced by Ince. However, as we shall 
see in Sec.\ \ref{q4sec}, both pairs of independent functions are relevant 
in connection with period-doubling bifurcations. The Lam\'e functions found
for $0\leq p\leq 3$ are explicitly given in \tab{q4lame2} of Sec.\ 
\ref{q4sec}. 

\section{The H\'enon-Heiles potential}
\label{hhsec}

\noindent
We investigate here the role of the straight-line librating orbit A in 
the H\'enon-Heiles (HH) potential \cite{hh}. The Hamiltonian reads in 
scaled coordinates
\begin{equation}
e = 6H = 6\left[\frac12\,(p_x^2+p_y^2) + V_{\rm H\!H}(x,y)\right], 
   \qquad V_{\rm H\!H}(x,y) = \frac12\,(x^2+y^2) + x^2y- \frac13\,y^3,
\label{hhxy}
\end{equation}
whereby the scaled energy is $e=1$ at the saddle points. The Newton 
equations of motion are
\begin{eqnarray}
\ddot x + (1 + 2y)\, x & = & 0\,, \label{hheomx}\\
\ddot y + y - y^2 + x^2& = & 0\, . \label{hheomy}
\end{eqnarray}
These equations, and therefore the classical dynamics of the HH potential, 
depend only on the scaled energy $e$ as a single parameter. In our 
numerical investigations we have solved Eqs.\ (\ref{hheomx},\ref{hheomy}) 
and determined the periodic orbits by a Newton-Raphson iteration using 
their stability matrix \cite{kabr}.

The basic periodic orbits with shortest periods found in the HH potential 
have been discussed mathematically by Churchill \etal\ \cite{chur}; an 
exhaustive numerical search and classification has been performed by Davies 
\etal\ \cite{hhdb}. We focus here on the straight-line A orbit which exists 
along the three symmetry axes of the HH potential, one of which coincides 
with the $y$ axis. This orbit undergoes an infinite series of bifurcations 
which were studied in Ref.\ \cite{mbgu}. They form a geometric progression 
on the scaled energy axes $e$, cumulating at the critical energy $e=1$ where 
the period $\TA$ becomes infinity and the orbit A becomes non-compact. All 
the orbits bifurcated from it exist, however, also at $e>1$ and stay in a 
bounded region of the $(x,y)$ space. Vieira and Ozorio de Almeida \cite{vioz} 
have investigated some of these orbits at $e>1$, both numerically 
and semi-analytically using Moser's converging normal forms near a harmonic 
saddle. In Fig.\ \ref{selfsim} we show the shapes of the orbits born at the 
isochronous bifurcations of orbit A, i.e., of those orbits having the same 
period $\TA$ as orbit A at the bifurcation points. The subscripts of their 
names O$_\sigma$ indicate their Maslov indices $\sigma$ needed in the context 
of the semiclassical periodic orbit theory \cite{gutz,gubu,book}. Although we 
make no use of the Maslov indices in the present paper, they are a convenient 
means of classification of the bifurcated orbits, as will become evident from 
the systematics below.

All orbits shown in \fig{selfsim} are evaluated at the barrier energy $e=1$. 
In the upper part of the figure, the $x$ axis has been zoomed by a factor 
0.163 from each panel to the next, in order to bring the shapes to the same 
scale. The orbits look practically identical in the lower 97\% of their 
vertical range, but near the barrier ($y=1$) they make one more oscillation 
in the $x$ direction with each generation. In the lower part of the figure, 
we have zoomed also the $y$ axis by the same factor from one panel to the 
next and plotted the top part of each orbit, starting from $y=1$. In these 
blown-up 

\Figurebb{selfsim}{-5}{280}{560}{571}{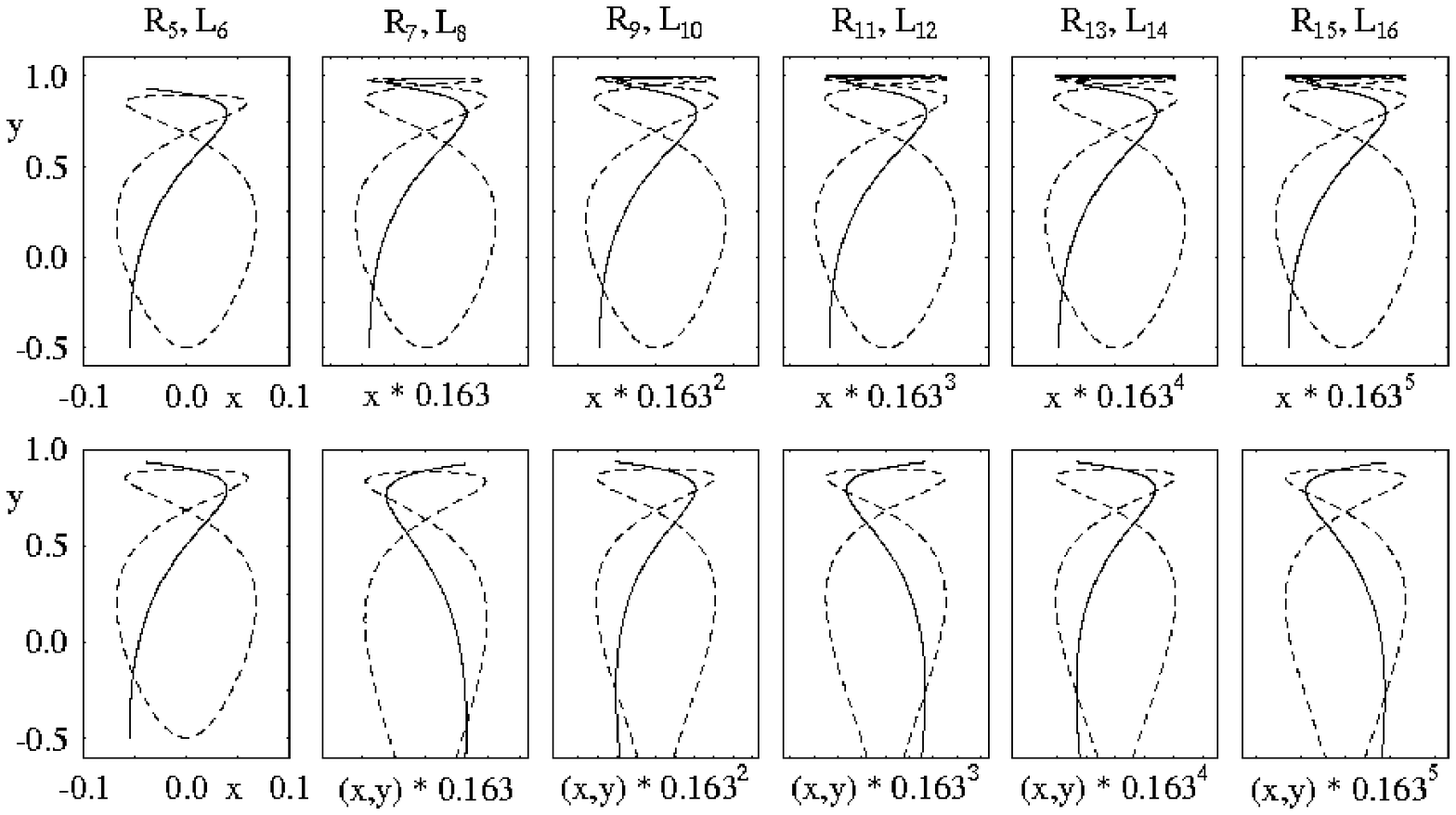}{7.9}{16.6}{
Orbits bifurcated from the A orbit in the HH potential, evaluated at 
energy $e=1$. The subscripts give their Maslov indices $\sigma$.
{\it Dashed lines:} rotations R$_\sigma$, {\it solid lines:} librations
L$_\sigma$ (only one libration orbit is shown for each index; its 
partner is obtained by reflection at the vertical symmetry line 
containing orbit A). {\it Top panels:} successive scaling of $x$ axis 
from left to right with the factor 0.163. {\it Bottom panels:} 
successive scaling of both axes with the same factor; along
the $y$ axis only the top part (starting from $y=1$) is shown.
}

\noindent
scales, the tips of the orbits exhibit a perfect self-similarity 
which has been described by analytical scaling constants in Ref.\ 
\cite{mbgu}.

In Fig.\ \ref{zoom} we show the stability discriminant $\trM$ for the orbit 
A (with its Maslov index $\sigma$ increasing by one unit at each bifurcation) 
and the orbits born at its isochronous bifurcations, plotted versus scaled 
energy $e$. In the lowest panel, we see the uppermost 3\% of the energy 
scale available for the orbit A. The first bifurcation occurs at $e_5 = 
0.969309$, where A$_5$
becomes unstable (with $\trMA$ $>2$) and the stable orbit R$_5$ is born. At 
$e_6 = 0.986709$, orbit A$_6$ becomes stable again and a new unstable orbit 
L$_6$ is born. In the middle panel, we have zoomed the uppermost 3\% of the 
previous energy scale. Here the behavior of A repeats itself, with the new 
orbits R$_7$ and L$_8$ born at the next two bifurcations. Zooming with the 
same factor to the top panel, we see the birth of R$_9$ and L$_{10}$. This 
can be repeated {\it ad infinitum}: each new figure will be a replica of the 
previous one, with all the Maslov indices increased by two units and with 
$\trMA$ oscillating forever. This fractal behaviour is characteristic of the
``Feigenbaum route to chaos'' \cite{feig,fei2,fei3}, although the present 
system is different from the H\'enon map in that the pitchfork bifurcations
seen in \fig{zoom} here are isochronous due to the reflection symmetry of the
HH potential around the lines on which the bifurcating orbit A is situated
(see Refs.\ \cite{nong,maod,then} for a discussion of the non-generic nature 
of the bifurcations in potentials with discrete symmetries). Also, the 
successive bifurcations happen here from one and the same orbit A, whereas 
in the standard Feigenbaum scenario one studies repeated period doubling 
bifurcations.

Note that the functions $\trM(e)$ of the bifurcated orbits in \fig{zoom} are
approximately linear and intersect at $e=1$ in two points, one for the 
librating orbits L$_\sigma$ with $\trM_{\rm L}(e=1)=+8.183$ (lying outside the 
figure), and one for the rotating orbits R$_\sigma$ with $\trM_{\rm R}(e=1)=
-4.183$. We shall derive this linear behaviour in the limit $e\rightarrow1$ 
from an asymptotic analytical evaluation of $\trMA$ in a forthcoming
publication \cite{fmmb}.

\Figurebb{zoom}{-60}{152}{560}{695}{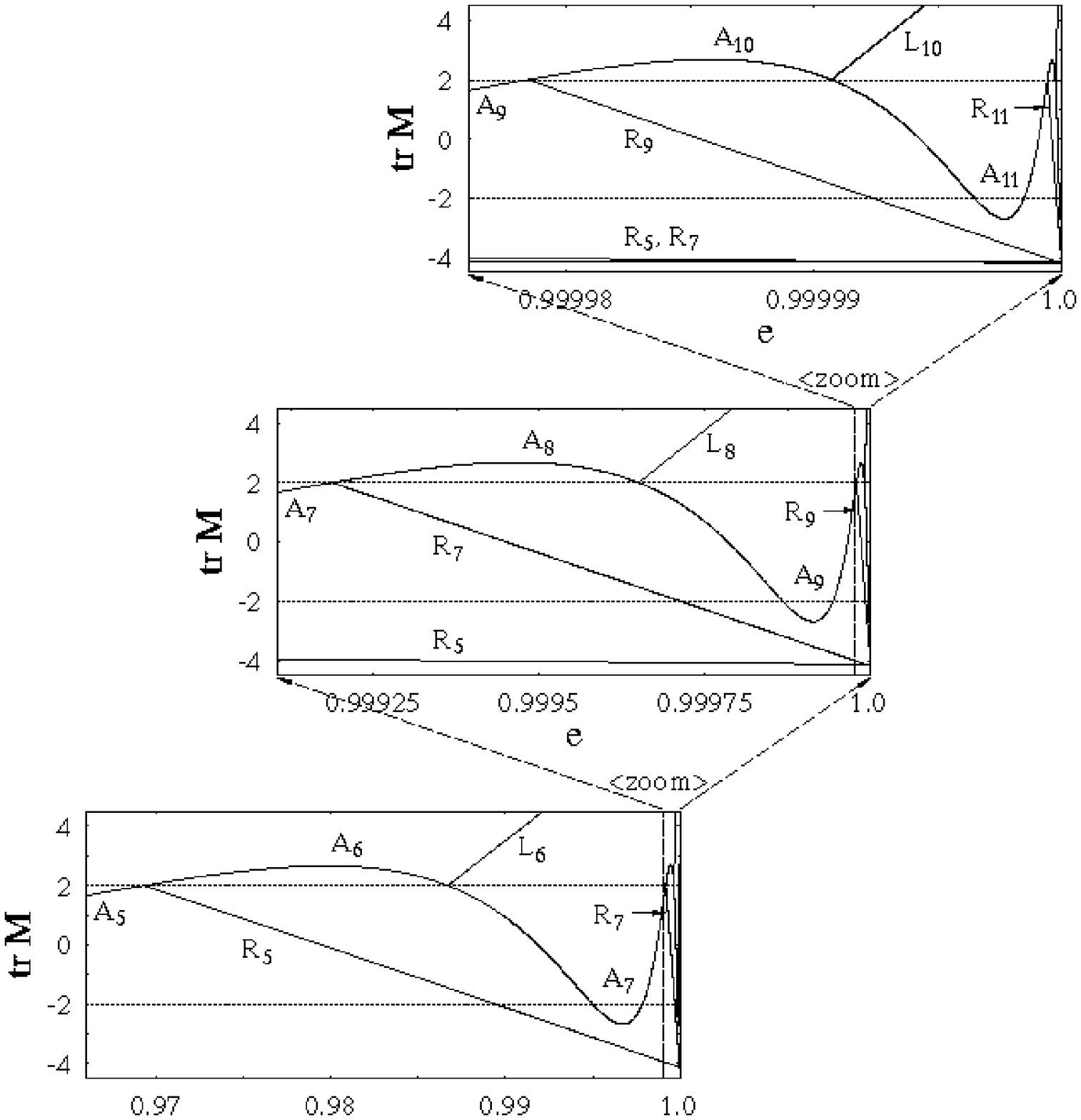}{12.2}{16.6}{
Trace of stability matrix M of orbit A and the orbits born at successive 
pitchfork bifurcations in the HH potential, plotted versus scaled energy $e$. 
{\it From bottom to top:} successively zoomed energy scale near $e=1$.
}

Presently we focus on the shapes of the orbits born at the bifurcations of A.
Infinitely close to the bifurcation points, their motion in the transverse
$x$ direction is given by periodic solutions of the stability equation
\eq{hill}. Note that this equation is identical with the full equation of 
motion \eq{hheomx} in the $x$ direction, which happens to be linear for the 
HH potential. The function $y_A(t)$ describing the A orbit can easily be 
found analytically \cite{mbgu} and is given, with the initial condition 
$y_A(0)=y_1$, by
\be
y_A(t) = y_1 + (y_2-y_1)\,\sn^2(z,k)\,,
\label{yaoft}
\ee
where $z$ is the scaled time variable
\be
z = \sqrt{(y_3-y_1)/6}\,t = at \,,
\label{hhta}
\ee
and $y_i$ ($i=1,$ 2, 3) are the roots of the cubic equation $e=6\,V_{\rm 
H\!H} (x=0,y) = 3\,y^2-2\,y^3$. $y_1$ and $y_2$ are the turning points 
of the A orbit, whose period is
\be
\TA=2\sqrt{6/(y_3-y_1)}\,\K = (2/a)\,\K\,.
\label{ta}
\ee
The modulus of the elliptic integral is given by
\be
   k^2 = (y_2-y_1)/(y_3-y_1)
\ee
and tends to unity for $e\rightarrow 1$ where $y_2=y_3$. Rewriting Eq.\ 
\eq{hheomx} in terms of the scaled time variable $z$, it becomes identical
with the Lam\'e equation \eq{lameq}, with
\be
h = 6\,(1+2\,y_1)/(y_3-y_1)\,, \qquad n(n+1) = -12 \quad 
    \Leftrightarrow \quad  n=-1/2\pm(i/2)\sqrt{47}.
\label{hnhh}
\ee

\Table{hhlame}{16.6}{
\begin{tabular}{|l|l|l|r|c||l|l|l|r|c|}
\hline
$e^{\star}_\sigma$ & O$_\sigma$ & $x_\sigma(t)$ \ & $r_{max}$ & $P$ &
$e_\sigma$ & O$_\sigma$ & $x_\sigma(t)$ \ & $r_{max}$ & $P$  \\
\hline
0.811715516     & F$_9$     & Ec$_n^3(at)$      & 13    & 4\K &
0.9693090904    & R$_5$     & Es$_n^4(at)$      & 20    & 2\K \\
0.915214692     & F$_{10}$  & Es$_n^3(at)$      & 12    & 4\K &
0.9867092353    & L$_6$     & Ec$_n^4(at)$      & 26   & 2\K \\
0.995013        & F$_{13}$  & Ec$_n^5(at)$      & 25   & 4\K &
0.9991878410    & R$_7$     & Es$_n^6(at)$      & 39   & 2\K \\
0.99784905      & F$_{14}$  & Es$_n^5(at)$      & 30   & 4\K &
0.9996498       & L$_8$     & Ec$_n^6(at)$      & 40   & 2\K \\
0.99986763      & F$_{17}$  & Ec$_n^7(at)$      & 53   & 4\K &
0.999978390     & R$_9$     & Es$_n^8(at)$      & 67   & 2\K \\
0.99994292      & F$_{18}$  & Es$_n^7(at)$      & 62   & 4\K &
0.9999906955    & L$_{10}$  & Ec$_n^8(at)$      & 104   & 2\K \\
0.999996482     & F$_{21}$  & Ec$_n^9(at)$      & 123   & 4\K & 
0.999999424     & R$_{11}$  & Es$_n^{10}(at)$   & 152  & 2\K \\
0.999998483     & F$_{22}$  & Es$_n^9(at)$      & 162  & 4\K & 
0.9999997525    & L$_{12}$  & Ec$_n^{10}(at)$   & 211  & 2\K \\
0.9999999065    & F$_{25}$  & Ec$_n^{11}(at)$   & 290  & 4\K & 
0.99999998475   & R$_{13}$  & Es$_n^{12}(at)$   & 450  & 2\K \\
0.9999999593    & F$_{26}$  & Es$_n^{11}(at)$   & 276  & 4\K & 
0.99999999343   & L$_{14}$  & Ec$_n^{12}(at)$   & 517  & 2\K \\
0.999999997514  & F$_{29}$  & Ec$_n^{13}(at)$   & 765  & 4\K & 
0.9999999996046 & R$_{15}$  & Es$_n^{14}(at)$   & 757  & 2\K \\
0.999999998928  & F$_{30}$  & Es$_n^{13}(at)$   & 890  & 4\K & 
0.9999999998249 & L$_{16}$  & Ec$_n^{14}(at)$   & 1203 & 2\K \\
\hline
\end{tabular}
\vspace*{-0.4cm}
}{~Bifurcation energies $e_\sigma$, $e^{\star}_\sigma$ of the A orbit in the 
H\'enon-Heiles (HH) potential, names O$_\sigma$ of the bifurcated orbits, 
Lam\'e functions of their motion $x_\sigma(t)$; cut-off order $r_{max}$ of 
the Fourier expansion, and period $P$ of the Lam\'e functions. The constant
$a$ is given in \eq{hhta}. {\it Right part:} isochronous bifurcations; {\it 
left part:} period-doubling bifurcations.}

\vspace*{-0.5cm}

Note that $h$ depends on the energy $e$. Hence, the discrete eigenvalues
$h=a_n^m,$ $b_n^m$ can be directly related to the bifurcation energies
$e_\sigma$ of the orbit A, and the corresponding Lam\'e functions Ec$_n^m(z)$ 
and Es$_n^m(z)$ to the motion $x(z)=x(at)$ of the new periodic orbits 
O$_\sigma$ born at the bifurcations. In \tab{hhlame} we give the bifurcation 
energies $e_\sigma$ obtained from the numerical computation of $\trMA$, the 
names O$_\sigma$ of the bifurcated orbits, and the corresponding Lam\'e 
functions with their periods in the variable $z=at$ given in \eq{hhta}. We 
also give the values $r_{max}$ at which the Fourier series \eq{ece} -- 
\eq{eso} have been truncated. The right part of the table contains the lowest
isochronous bifurcations seen in \fig{zoom} and the bifurcated orbits shown 
in \fig{selfsim}; their Lam\'e functions all have the period 2$\K$. 

The left part of \tab{hhlame} contains the lowest non-trivial 
period-doubling bifurcations (where $\trMA=-2$), which are also of 
pitchfork type, and the names of the orbits born thereby. Their Lam\'e 
functions have the period 4$\K$. To avoid ambiguities, we denote their 
bifurcation energies by $e^{\star}_\sigma$. The period-doubling 
bifurcations of new orbits with the Maslov indices 11, 12, 15, 16, etc., 
are trivial in the sense that they just involve the second iterates of 
orbit A and of the bifurcated orbits R$_5$, L$_6$, R$_7$, L$_8$, etc. 
The shapes of the first six non-trivial new orbits born at these
bifurcations are shown in \fig{hdoub}; they have similar scaling 
properties as those shown in \fig{selfsim}.

\Figurebb{hdoub}{20}{45}{852}{310}{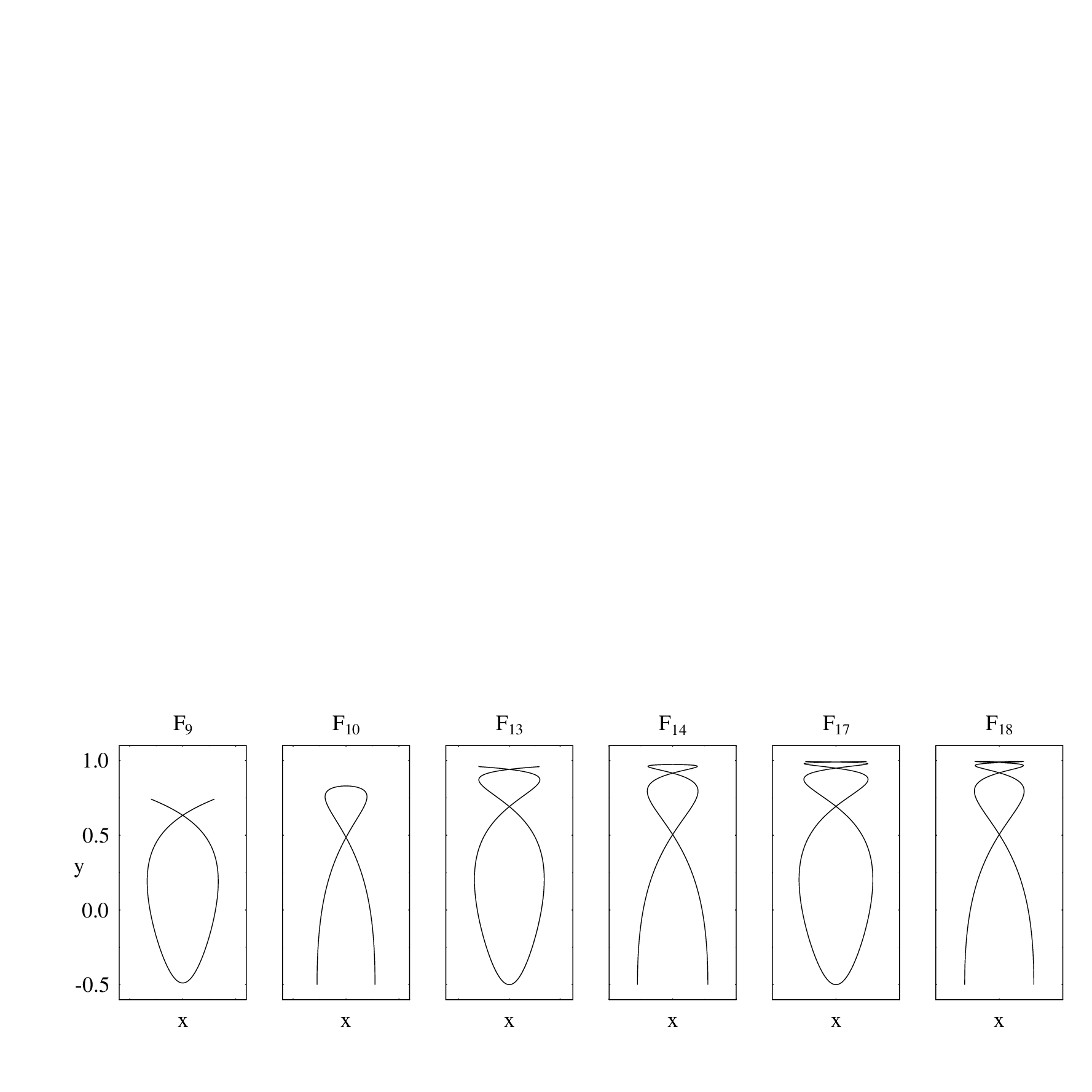}{5}{16.6}{
The periodic orbits born at the six lowest period-doubling bifurcations of 
orbit A in the H\'enon-Heiles potential. The $x$ axis has been scaled like 
in \fig{selfsim}.
}

Mathematically speaking, the periodic solutions for $x_\sigma(t)$ in the 
form of the Lam\'e functions exist only at the bifurcation energies 
$e_\sigma$. However, the bifurcated orbits exist for all $e\geq e_\sigma$. 
As long as the amplitude of their $x$ motion remains small, it must be
given by the Lam\'e equation \eq{lameq} with the constants appearing in
\eq{hnhh}. But this equation has only periodic solutions when $h$ has an
eigenvalue corresponding to a bifurcation energy $e_\sigma$. Therefore,
the bifurcated orbits must keep their $y$ motion ``frozen'' at $y(t) =
y_A(t)$ with the parameters corresponding to $e_\sigma$. Consequently, 
they also keep their
periods at the bifurcation values. This has been confirmed numerically,
as noticed already in \cite{mbgu}, to hold up to $e=1$ and even beyond. 
Within the same small-amplitude limit of the $x$ motion, the energy of 
the $y$ motion is frozen at its value $e_\sigma$, and the excess energy 
$e-e_\sigma$ is consumed to rescale the amplitude of $x(t)$. In other 
words, we can determine the normalization of the Lam\'e function 
$x_\sigma(t)$ of each bifurcated orbit by exploiting the energy 
conservation. This is most easily done at the time $t_0=T_A/2$ where $y(t)$ 
has its maximum value, i.e., $y(t_0)=y_2$, ${\dot y}(t_0)=0$, and around
which we know the symmetry of the Lam\'e functions. For the even functions 
Ec$_n^m$ we have ${\dot x}_\sigma(t_0)=0$, and $x_\sigma(t_0)$ is, with
\eq{hhxy}, found to be
\be
x_\sigma(t_0) = \sqrt{(e-e_\sigma)/3(1+2y_2)}\,.
\label{x0}
\ee 
For the odd functions Es$_n^m$ we have $x_\sigma(t_0)=0$, and their slopes
at $t_0$ are given by
\be
{\dot x}_\sigma(t_0) = \sqrt{(e-e_\sigma)/3}\,.
\label{xd0}
\ee 
In this way we can not only normalize the Lam\'e functions near the 
bifurcation points, but also predict their evolution at higher energies. 

In Figs.\ \ref{hhorb5} - \ref{hhorb14} show some of the periodic orbits 
obtained numerically from solving the equations of motion \eqq{hheomx}
{hheomy} at $e=1$ by solid lines, and compare them to those predicted in 
the frozen-$y$-motion approximation, using $y(t)= y_A(t)$ (given at their 
bifurcation energies $e_\sigma$ or $e^\star_\sigma$) and using for $x(t)$ 
the Lam\'e functions according to \tab{hhlame}, scaled as explained above. 
We see that in all cases, even for the lowest bifurcations, the new orbits 
keep their $y$ motion acquired at their bifurcation energies, up to $e=1$, 
indeed: the two curves $y(t)$ and $y_A(t)$ can hardly be distinguished.
As a consequence, 

\Figurebb{hhorb5}{-40}{60}{755}{565}{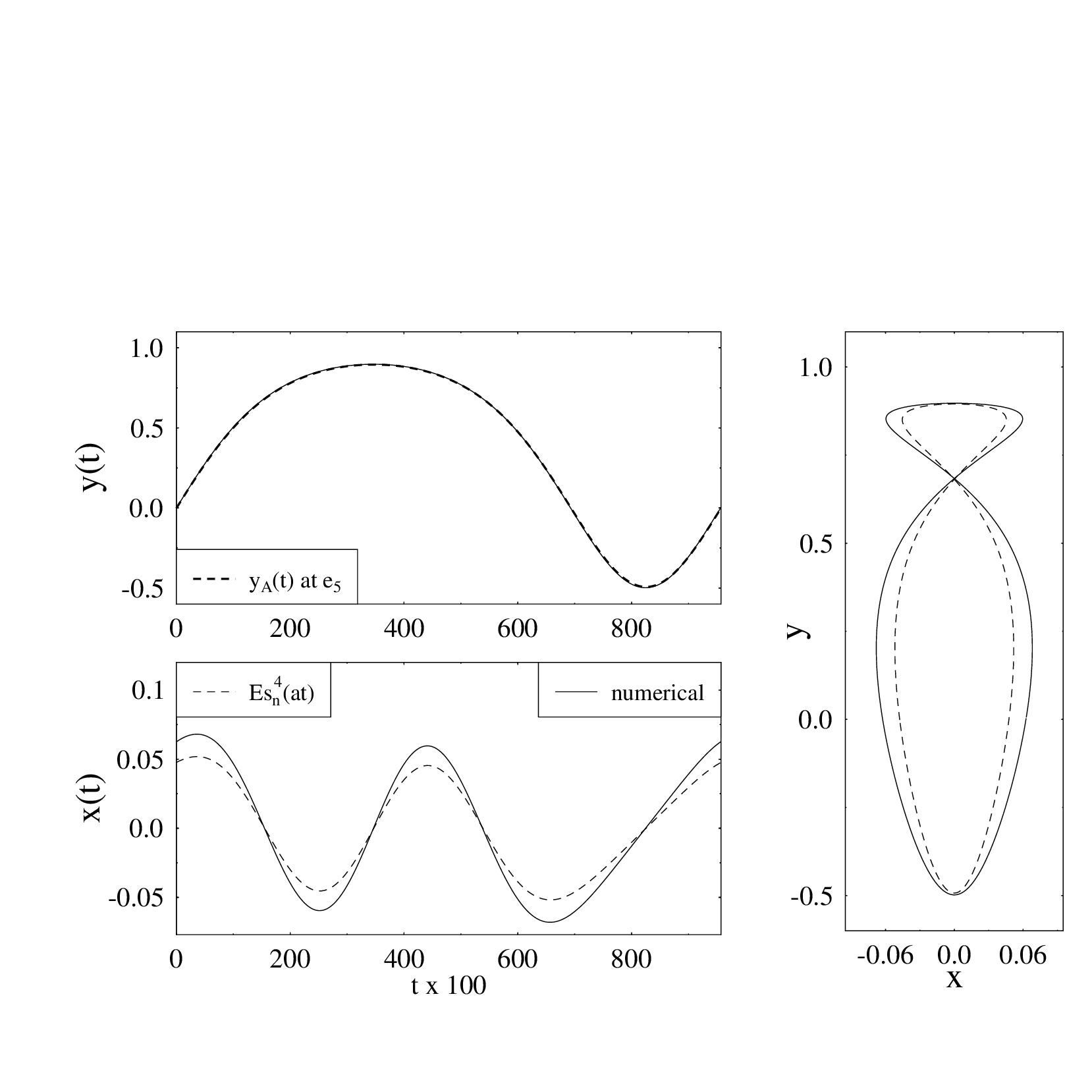}{8.5}{16.6}{
Orbit R$_5$ in the HH potential at $e=1$. {\it Left panels:} $y(t)$ and 
$x(t)$ versus time $t$ (in units of 0.01); {\it Right panel:} orbit in 
the $(x,y)$ plane. {\it Solid lines:} numerical results obtained by 
solving Eqs.\ (\ref{hheomx},\ref{hheomy}). {\it Dashed lines:} $y(t)$ 
given by $y_A(t)$ in \eq{yaoft} at the bifurcation energy $e_5$, and 
$x(t)$ given by the Lam\'e function according to \tab{hhlame}, scaled as 
described in the text.
}

\Figurebb{hhorbf13}{-20}{60}{755}{565}{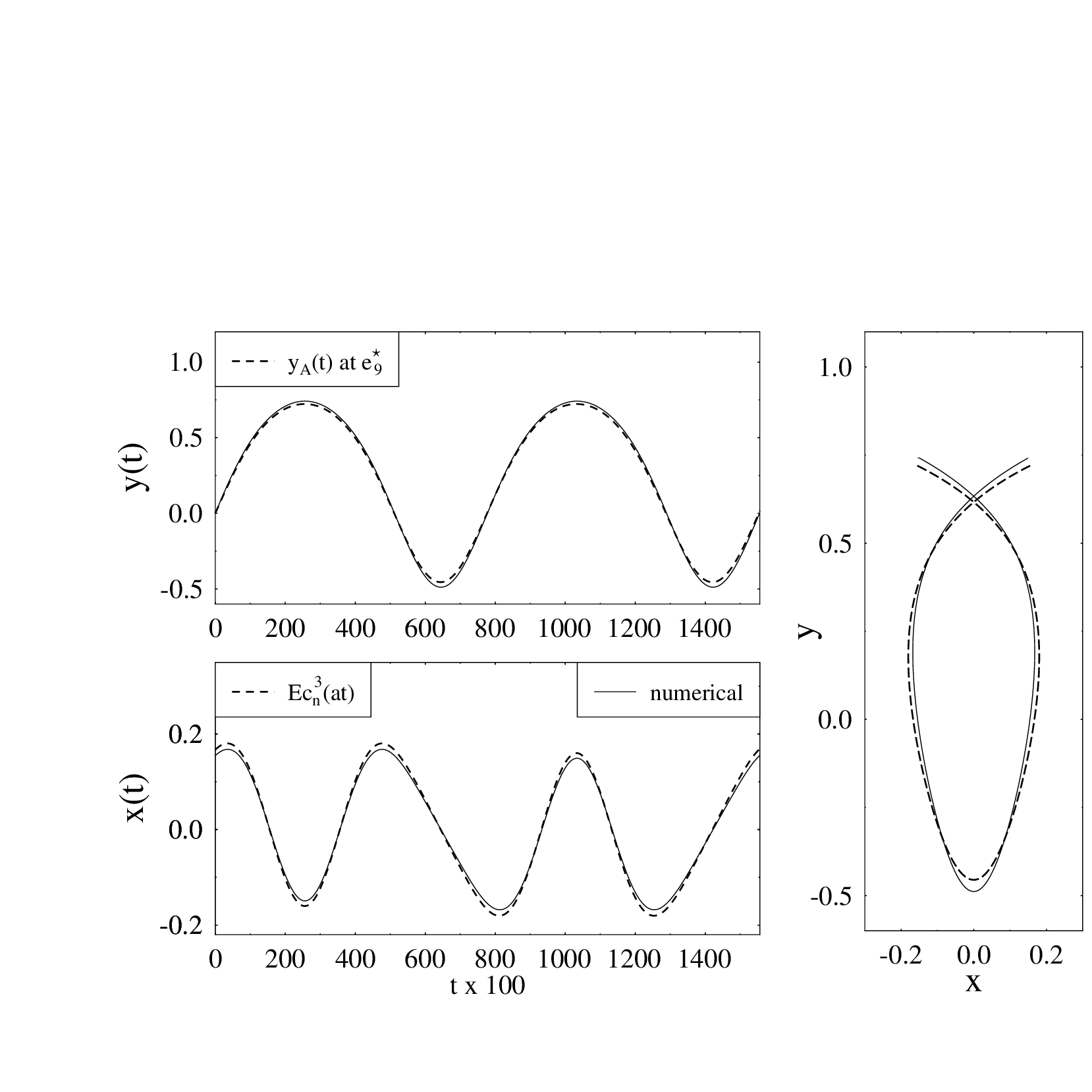}{8.5}{16.6}{
Same as \fig{hhorb5} for the orbit F$_{9}$. Note that this orbit has
twice the period of orbit A at $e^{\star}_{9}$ and is given by the
Lam\'e function Ec$_n^3(at)$ with period 4$\K$ and $m=3$ zeros in the
interval $0 \leq at < 2\K$.
}

\noindent
we can expect the functions $x(t)$ to be well described by the appropriate 
Lam\'e functions. This is, indeed, the case if the latter are correctly 
scaled. As we see, the normalization predicted by \eqq{x0}{xd0} is the 
better, the closer the bifurcation energy $e_\sigma$ comes to the 
saddle-point energy $e=1$. A rigorous justification of the 
frozen-$y$-motion approximation will be presented elsewhere \cite{fmmb}.

We point out that all orbits born at the isochronous pitchfork bifurcations 
in the HH system are given by Lam\'e functions with period 2$\K$, since the 
orbit A is given by $y_A(t)$ \eq{yaoft} and hence has the same period as the 
function appearing in the Lam\'e equation. The Lam\'e functions with period
4$\K$ must therefore correspond to orbits born at period-doubling 
bifurcations (see \tab{hhlame} and \fig{hhorbf13}). 

\Figurebb{hhorb8}{-40}{60}{755}{565}{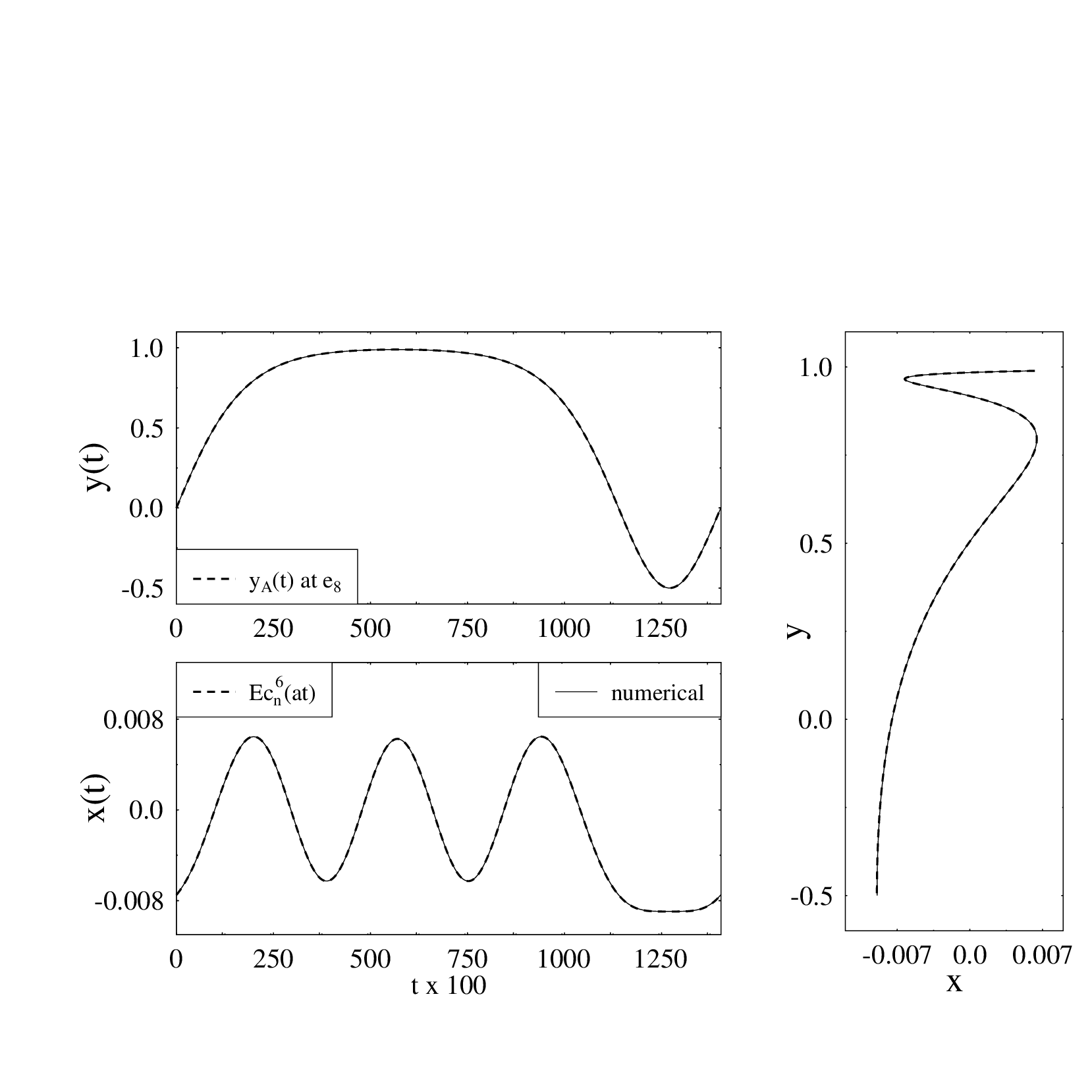}{8.5}{16.6}{
Same as \fig{hhorb5} for the orbit L$_8$.
}

\Figurebb{hhorb14}{-40}{60}{755}{565}{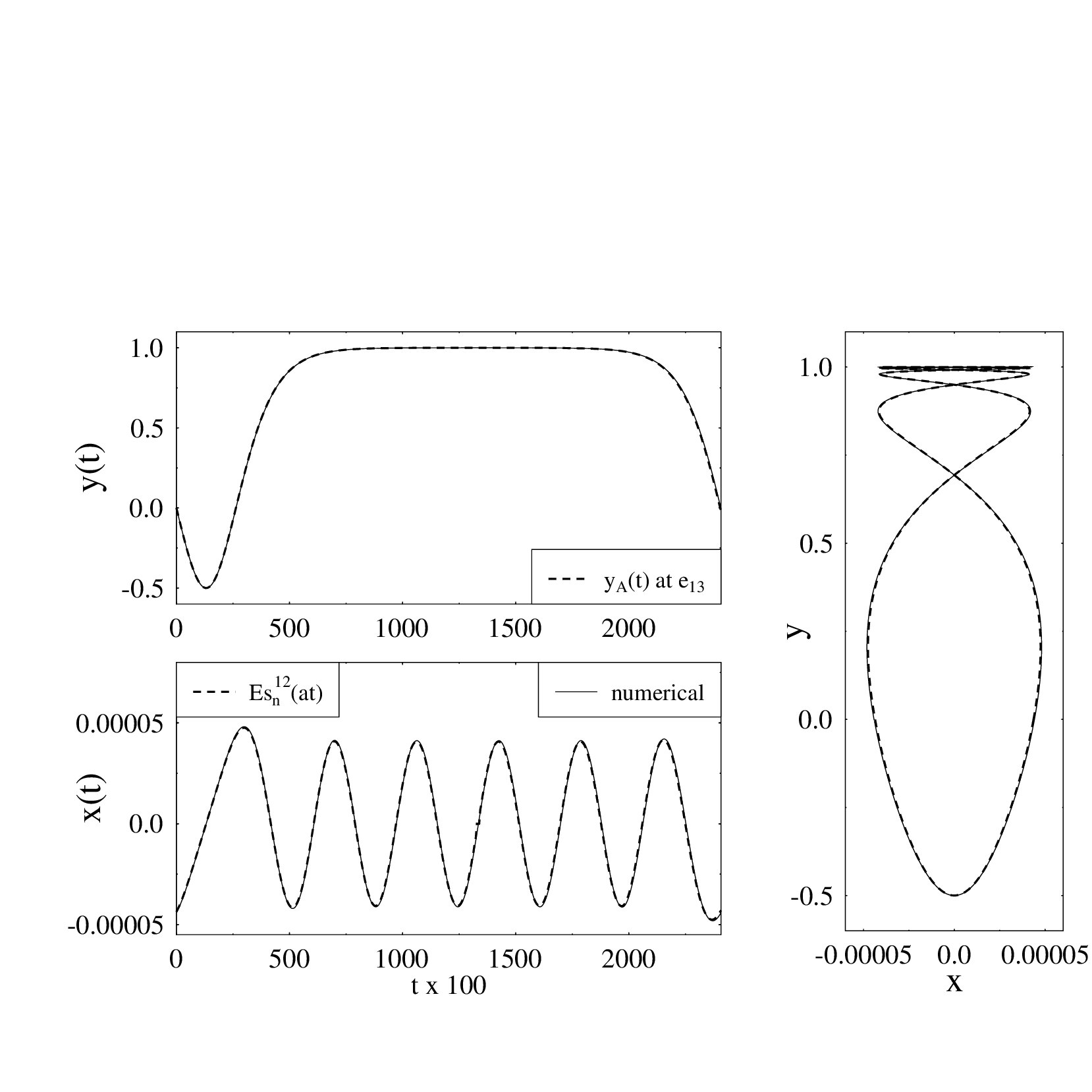}{8.5}{16.6}{
Same as \fig{hhorb5} for the orbit R$_{13}$. Note that $x(t)$ is harmonic 
over a long time while $y$ is close to the saddle ($y\simeq 1$); the period
of this harmonic oscillation is $\omega_\perp=2\pi/\sqrt{3}$, as shown
in Ref.\ \cite{mbgu} and also obtained from an asymptotic expansion of the
Lam\'e functions \cite{fmmb}.
}

Note also that according to bifurcation theory \cite{nong,maod,then,ssun}, 
two degenerate periodic orbits should be born at each isochronous pitchfork 
bifurcation. The librating orbits L$_\sigma$ come, indeed, in pairs that 
are symmetric to the $y$ axis, whereas the rotating orbits R$_\sigma$ can 
be run through in two opposite directions. Each of these pairs of orbits 
are, however, described by one and the same Lam\'e function for $x(t)$ 
which is invariant under the corresponding symmetry operation. This is in 
agreement with a theorem, proved by Ince \cite{inc2}, that there cannot 
exist two linearly independent periodic solutions of the Lam\'e equation 
to the same characteristic value of $h$. An exception of this theorem is 
given by transcendental and algebraic Lam\'e functions with integer or 
half-integer values of $n$ (see Sec.\ \ref{q4sec} for an example). 

We finally note that in the figures \ref{hhorb5} - \ref{hhorb14}, the 
time scales are not always normalized such that $t=0$ corresponds to 
$y(0)=y_1$, as assumed in Eq.\ \eq{yaoft}, but obtained rather randomly 
due to the way in which the periodic orbits where searched and found 
numerically. However, if we shift the time origin to $t'=0$ according to 
\eq{yaoft}, all figures illustrate how the Lam\'e functions Ec$_n^m(z)$ 
are even and the Es$_n^m(z)$ odd, according to their definition 
\cite{inc1,erd2}, around $t'=\K/a$ where $y(t')$ has its maximum. This 
demonstrates that the association of Lam\'e functions to bifurcated 
orbits allows one to understand (or predict) their symmetries.

\section{The quartic H\'enon-Heiles potential}
\label{h4sec}
 
We next investigate the quartic H\'enon-Heiles (H4) potential 
\cite{pert,hhun} with the scaled Hamiltonian
\be
e = 4H = 4\left[\frac12\,(p_x^2+p_y^2) + V_{\rm H4}(x,y)\right], \quad
V_{\rm H4}(x,y) = \frac12\,(x^2+y^2) - \frac14\,(x^4+y^4) + \frac32\,x^2y^2,
\label{r4xy}
\ee
which is similar to the HH potential but has four saddles at the scaled
energy $e=1$; it has reflection symmetry at both coordinate axes and both 
diagonals. It contains straight-line librating orbits along all four
symmetry lines; two of them, which we call again orbits A, oscillate between 
the two saddles lying on the coordinate axes. To be specific, we choose 
again the A orbit along the $y$ axis. It has the same behaviour as the A
orbit in the HH potential, but it approaches a saddle at both ends. Its
motion is, for $y_A(0)=0$, given by
\be
y_A(t) = y_1\,\sn(at,k)\,, \qquad a=y_2/\sqrt{2}\,, 
                           \qquad k = y_1/y_2\,,
\label{yar4}
\ee
and its period is
\be
\TA = 4\sqrt{2}\K/y_2 = 4\K/a\,.
\label{perar4}
\ee
Hereby $\pm y_1$ and $\pm y_2$ are the solutions of $e=4V_{\rm H4}(x=0,y)
= 2\,y^2 -y^4$, i.e.,
\be
y_1=\sqrt{1-\sqrt{1-e}}\,, \qquad y_2=\sqrt{1+\sqrt{1-e}}\,,
\ee
and $\pm y_1$ are the turning points of the orbit.

The linearized equation of motion in the $x$ direction, which decides about
the stability of the orbit A, is for the H4 potential
\be
{\ddot x}(t) + [1+3y^2(t)]\,x(t) = 0\,,
\ee
neglecting here explicitly a term of order $x^3$. Inserting the solution
for $y_A(t)$ in \eq{yar4} and transforming to the scaled time variable
$z=at$ leads again to the Lam\'e equation \eq{lameq} with
\be
h = 2/y_2^2\,, \qquad n(n+1) = -6 \quad 
    \Leftrightarrow \quad  n=-1/2\pm(i/2)\sqrt{23}\,.
\ee
Compared to the HH potential, we have now a new situation which is a 
consequence of the higher symmetry of the H4 potential: the periodic
function $\sn^2(z,k)$ appearing in the stability equation has {\it half}
the period, namely 2$\K$, of that of the orbit A itself \eq{perar4}.
Therefore, all its periodic solutions with period $2\K$, corresponding
to Lam\'e functions with an even number $m$ of zeros, also have the period 
$\TA/2$ at the bifurcations. The solutions involving the Lam\'e functions
with odd $m$ share their periods $4\K=a\TA$ with that of the A orbit. 

The systematics of the isochronous bifurcations of the A orbit for 
increasing bifurcation energies $e_\sigma$ is given in \tab{r4lame}. The
new orbits appear with $m=2,$ 3, 4, \dots, with alternatingly odd and 
even Lam\'e functions. Like in the HH case, the Ec$_n^m$ correspond to 
librations L$_\sigma$ and the Es$_n^m$ to rotations R$_\sigma$. They 
appear alternatingly as 2$\K$ and 4$\K$ periodic functions. The orbits 
given in the left part of the table are born stable and remain stable up
to $e>1$, whereas those in the right part are unstable at all energies.

\Table{r4lame}{16.6}{
\begin{tabular}{|l|l|l|r|c||l|l|l|r|c|}
\hline
$e_\sigma$ & O$_\sigma$ & $x(t)$ & $r_{max}$ & $P$ & 
$e_\sigma$ & O$_\sigma$ & $x(t)$ & $r_{max}$ & $P$ \\
\hline
0.8561220          & R$_5$     & Es$_n^2(at)$   & 7   & 2\K &
0.8967139          & L$_6$     & Ec$_n^2(at)$   & 9   & 2\K \\
0.9841765          & L$_7$     & Ec$_n^3(at)$   & 11  & 4\K &
0.9889128          & R$_8$     & Es$_n^3(at)$   & 12  & 4\K \\
0.9982845          & R$_9$     & Es$_n^4(at)$   & 18  & 2\K &
0.9988004          & L$_{10}$  & Ec$_n^4(at)$   & 22  & 2\K \\
0.9998140          & L$_{11}$  & Ec$_n^5(at)$   & 29  & 4\K &
0.99986995         & R$_{12}$  & Es$_n^5(at)$   & 32  & 4\K \\
0.99997983         & R$_{13}$  & Es$_n^6(at)$   & 46  & 2\K &
0.99998590         & L$_{14}$  & Ec$_n^6(at)$   & 48  & 2\K \\
0.999997812        & L$_{15}$  & Ec$_n^7(at)$   & 77  & 4\K &
0.9999984705       & R$_{16}$  & Es$_n^7(at)$   & 98  & 4\K \\
0.9999997627       & R$_{17}$  & Es$_n^8(at)$   & 117 & 2\K &
0.9999998340       & L$_{18}$  & Ec$_n^8(at)$   & 134 & 2\K \\
0.9999999742       & L$_{19}$  & Ec$_n^9(at)$   & 194 & 4\K &
0.9999999820       & R$_{20}$  & Es$_n^9(at)$   & 236 & 4\K \\
\hline
\end{tabular}
}{~The same as in \tab{hhlame}, but only for the isochronous bifurcations
of the A orbit in the quartic H\'enon-Heiles (H4) potential. Note the
alternating appearance of 2$\K$ and 4$\K$ periodic Lam\'e functions with
increasing bifurcation energy $e_\sigma$. The constant $a$ is given in
Eq.\ \eq{yar4}. Orbits appearing on the left side are stable up to $e>1$,
those on the right side are unstable at all energies.}

\Figurebb{r4trme}{20}{283}{559}{559}{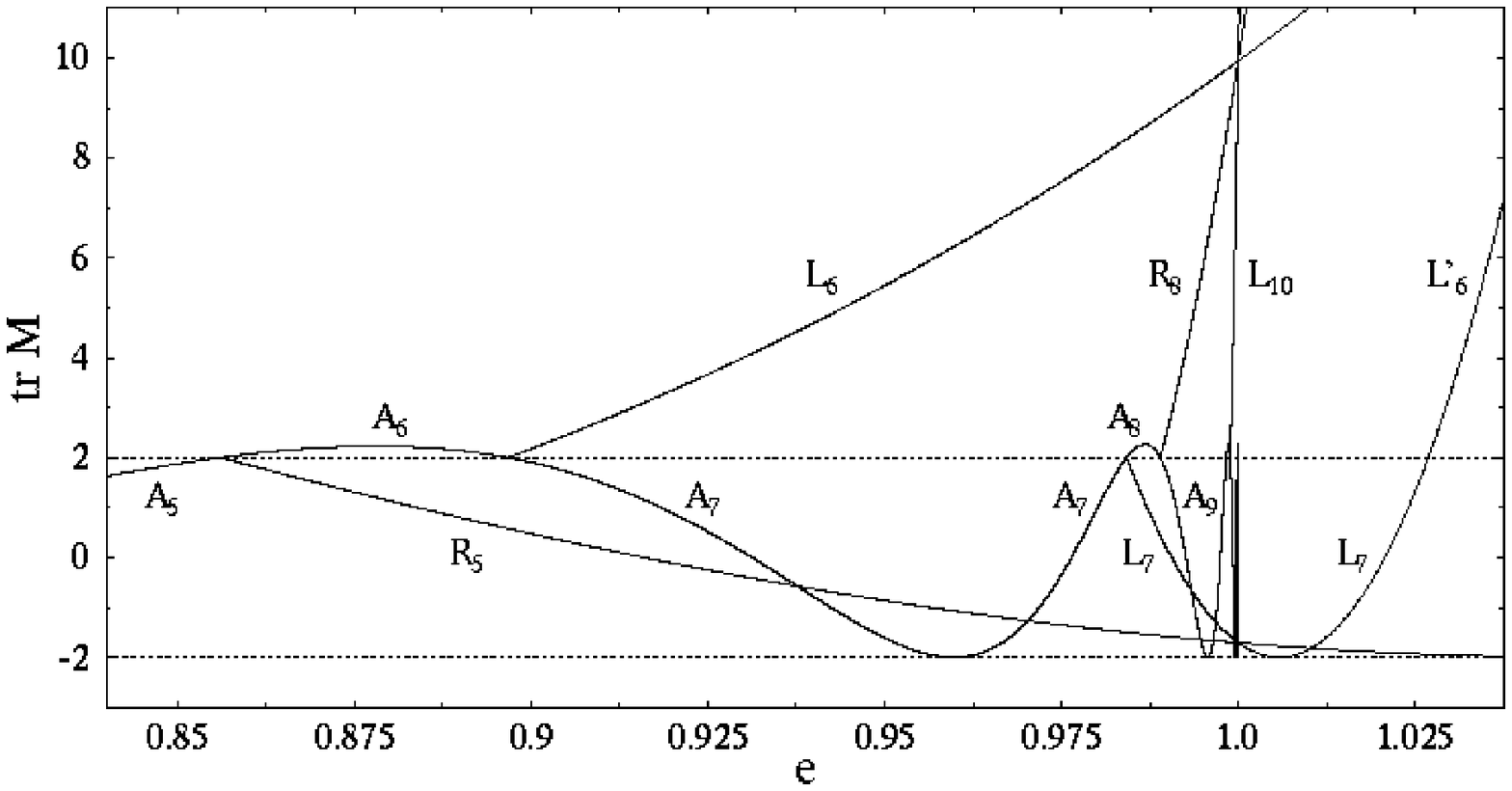}{8}{16.6}{
Stability discriminant of the shortest orbits in the H4 potential, plotted
versus scaled energy $e$ (cf.\ \tab{r4lame} for the bifurcation energies). 
L$_7$ bifurcates at $e\sim 1.027$ and becomes L$'_6$.
}

The non-trivial period-doubling bifurcations in this potential are of
island-chain type, and the orbits born thereby are given by Lam\'e
functions of period 8$\K$. We will not investigate them here, but refer
to the analogous situation in the quartic oscillator potential discussed
in Sec.\ \ref{q4sec}.

In \fig{r4trme} we show the stability discriminant $\trM$ versus energy 
$e$ for the orbit A and the orbits born at its lowest isochronous 
bifurcations. Different from the HH potential, here the functions 
$\trM(e)$ of the bifurcated orbits are, to a good approximation, 
quadratic in $e$. This can be derived analytically \cite{fmmb}. It is 
also striking that, different from \fig{zoom}, $\trM$ of all the orbits 
shown is always larger than or equal to $-2$. This behaviour, together 
with the systematics seen in \tab{r4lame}, can be explained by the 
following arguments.

Bearing in mind that the stability matrix of the second iterate O$^2$ of a 
periodic orbit O is just M$_{\rm O}^2$, where M$_{\rm O}$ is that of the 
primitive orbit, one easily finds that its discriminant is
\be
\trM_{{\rm O}^2} = \trM^2_{\rm O} = (\trM_{\rm O})^2 - 2\,,
\label{trmosq}
\ee

\noindent
which can never be less than $-2$. Hence, we can mathematically understand
$\trM$ of the orbit A in \fig{r4trme} to be that of a second iterate. Its
primitive is half of the orbit A, having the same period as that of the
function $\sn^2$ in the Lam\'e equation for its stability, and having a
discriminant $\trM$ which oscillates around zero, exceeding the values $+2$ 
and $-2$ on both sides symmetrically, like $\trMA$ in the HH potential
(\fig{zoom}). 
Their second iterates, which correspond to the full 
bifurcated orbits, therefore have a discriminant $\trM$ which is quadratic 
in $e$. These features, together with the systematics in \tab{r4lame}, are 
all a consequence of the C$_{4v}$ symmetry of the H4 potential, including 
the reflection symmetry at the $x$ axis that divides the A orbit (and all
the bifurcated orbits discussed here) into to equal halves. The quadratic
behaviour of $\trM$ of the bifurcated orbits is also consistent with the
fact that their next period-doubling bifurcations (where $\trM=-2$) are
symmetry breaking (see Ref.\ \cite{then} for details).   

Like in the HH case, the values of $\trM$ of the bifurcated orbits 
intersect in two points at $e=1$, one with $\trM(e=1)=-1.711$ for the 
orbits born stable, and one with $\trM(e=1) =+9.991$ for the orbits born 
unstable.

We thus obtain the result that Lam\'e functions with both periods 2$\K$
and 4$\K$ describe the orbits born at isochronous bifurcations of the A
orbit. Two examples are shown in Figs.\ \ref{r4orb14} and \ref{r4orb16}, 
where the $x$ motion of the orbits L$_{14}$ and R$_{16}$ is given by the 
functions Ec$_n^6(z)$ and Es$_n^7(z)$, respectively. 

\Figurebb{r4orb14}{-40}{60}{755}{565}{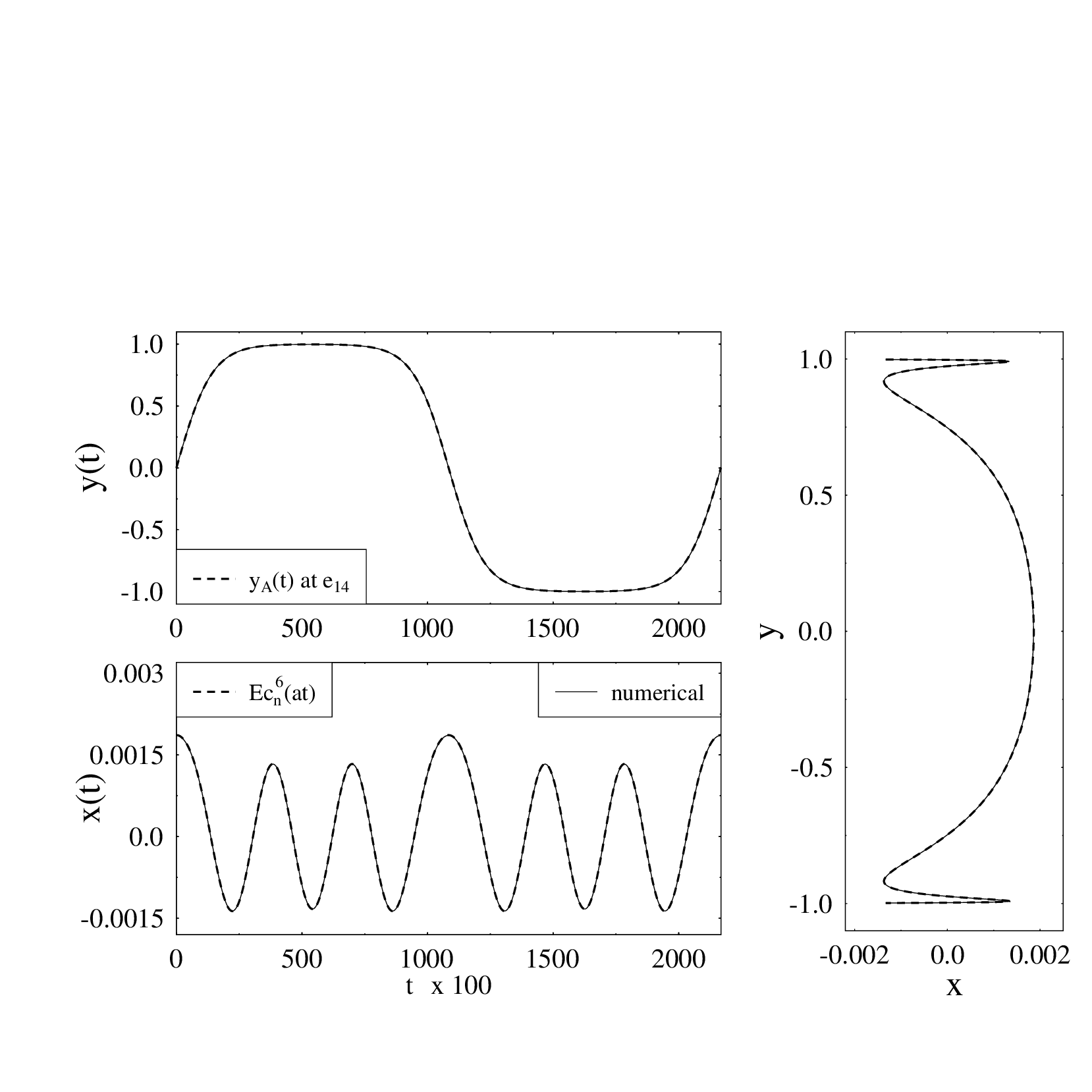}{8.5}{16.6}{
Same as \fig{hhorb5} for the orbit L$_{14}$ in the H4 potential (cf.\
\tab{r4lame}). Note that $x(t)$ has half the period of $y(t)$.
}

\noindent
Whereas the latter shares its period with orbit A, the former has half 
its period. Like before, these orbits are evaluated at the critical 
energy $e=1$. The normalization of the Lam\'e functions has been chosen 
as for the HH potential, using the frozen-$y$-motion approximation and
energy conservation, leading here to
\be
x_\sigma(t_0) = \sqrt{(e-e_\sigma)/2(1+3y_1)}\, \qquad
{\dot x}_\sigma(t_0) = \sqrt{(e-e_\sigma)/2}\,.
\ee

\Figurebb{r4orb16}{-40}{60}{755}{565}{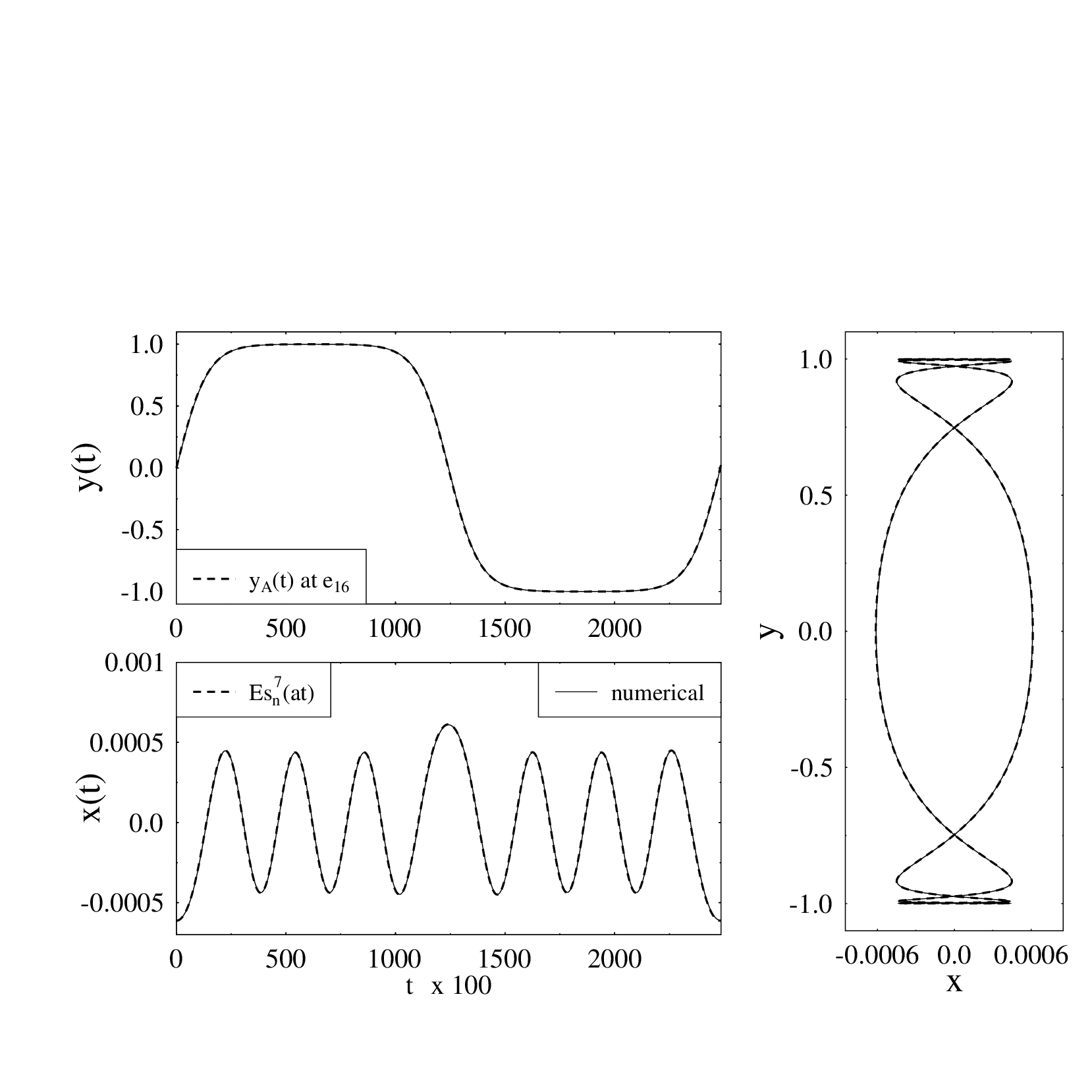}{8.5}{16.6}{
Same as \fig{hhorb5} for the orbit R$_{16}$ in the H4 potential (cf.\
\tab{r4lame}).
}

\vfill
\newpage

\section{The homogeneous quartic oscillator}
\label{q4sec}

We now turn to the quartic oscillator (Q4) potential
\be
V_{Q4}(x,y)= \frac14\left(x^4+y^4\right) + \frac{\epsilon}{2}\, x^2 y^2 
\label{q4xy}
\ee
which has been the object of several classical and semiclassical studies
\cite{q4po}. Since it is homogeneous in the coordinates, the Hamiltonian
can be rescaled together with coordinates and time such that its classical
mechanics is independent of energy. Consequently, the system parameter is
here not the energy but the parameter $\epsilon$. The potential \eq{q4xy}
has the same symmetry as the H4 potential \eq{r4xy} and, correspondingly,
possesses periodic straight-line orbits along both axes. The motion of the
A orbit along the $y$ axis is given by 
\be
y_A(t) = y_0\,\cn(y_0t,k)\,, \qquad y_0=(4E)^{1/4}, \qquad k^2=1/2
\label{yaq4}
\ee
with the period $T_A = 4\,\K/y_0$. Its turning points are $\pm y_0$. Note
that this solution does not depend on the value of $\epsilon$. The 
stability of the orbit A, however, does depend on $\epsilon$. The 
linearized equation of motion for the transverse $x$ motion yields, after 
transformation to the coordinate $z=y_0t$, the Hill equation
\be
x''(z) + \epsilon\,[1-\sn^2(z,k)]\,x(z) = 0\,.
\label{lameq4}
\ee
This is a special case of the Lam\'e equation with
\be
h = \epsilon = \frac12 \,n\,(n+1)\,,
\label{hn}
\ee
where we have used $k^2=1/2$. The nice feature is that we here know
analytically the eigenvalues $h=h_n$ of the Lam\'e equation, namely those 
given in Eq.\ \eq{hn}. This agrees with the analytical result for the 
stability discriminant of the A orbit \eq{yaq4}, which has been derived 
long ago by Yoshida \cite{yosh}:
\be
\trMA = 4\,\cos\left[\frac{\pi}{2}\sqrt{1+8\epsilon}\right] + 2\,.
\label{trmaq4}
\ee
It is easy to see that the bifurcation condition $\trMA=+2$ leads exactly 
to the values \eq{hn} of the parameter $\epsilon$.

The periodic solutions of the Lam\'e equation \eq{lameq4} at the 
bifurcation values of $h=\epsilon$ are the Lam\'e polynomials discussed 
already in Sec.\ \ref{lamsec}. Their explicit forms for $n=0,\,1,\,\dots,$ 
15 are given in \tab{q4lame}, using the short notation $\cn=\cn(z,k)$, 
etc., and a normalization such that their leading coefficient is unity. We 
also give in \tab{q4lame} the names of the new-born orbits, using the same 
nomenclature as for the H4 potential in the previous section. Their shapes 
are shown in \fig{plamorb}. They have exactly the same topologies as the 
orbits of the H4 potential and are again described by Lam\'e functions of 
pairwise alternating periods $2\K$ and $4\K$, as seen also in \tab{q4lame}.
Each of these orbits has a discrete degeneracy of 2, which is due to the 
time reversal symmetry for the rotations and to the reflection symmetries 
about the coordinate axes for the librations. 

A special comment is due concerning the cases $n=1$ and $n=2$ which
correspond to $\epsilon=1$ and 3, respectively. For these values of the 
parameter $\epsilon$, the Q4 potential is integrable \cite{q4po,yosh} 
and no bifurcations occur for the shortest orbits. Nevertheless, the 
Lam\'e equation \eq{lameq4} possesses mathematically the solutions
Ec$_1^1$ and Es$_2^1$, respectively. The orbits B$_3$ and C$_4$ given
in \tab{q4lame} and \fig{plamorb} have topologically the shapes given
by these Lam\'e polynomials, but it should be emphasized that they are
not generated through bifurcations but are generic orbits existing at all 
values of $\epsilon$. Under the symmetry operation $\epsilon\longrightarrow 
(3-\epsilon)/(1+\epsilon)$, which corresponds to a rotation about 45 
degrees and simultaneous stretching of the potential \cite{q4po}, the 
orbits of type A are mapped onto the 

\Table{q4lame}{16.6}{
\begin{tabular}{|r|r|l|l|c|}
\hline
$n$ & $\epsilon_n$ & O$_\sigma$ & Lam\'e polynomial for $x(t)$ & $P$ \\
\hline
0   & 0    & L$_3$     & Ec$_0^0\;$ = \,1                            & 2\K \\
1   & 1    & [B$_3$]   & [Ec$_1^1\;$ = \,cn]                         & 4\K \\
2   & 3    & [C$_4$]   & [Es$_2^1\;$ = \,dn\,\sn]                    & 4\K \\
3   & 6    & R$_5$     & Es$_3^2\;$ = \,cn\,dn\,\sn                  & 2\K \\
4   & 10   & L$_6$     & Ec$_4^2\;$ = \,$1 - \frac53\,\cn^4$         & 2\K \\
5   & 15   & L$_7$     & Ec$_5^3\;$ = \,cn\,($1-\frac75\,\cn^4$)     & 4\K \\
6   & 21   & R$_8$     & Es$_6^3\;$ = \,dn\,\sn\,($1-3\,\cn^4$)      & 4\K \\
7   & 28   & R$_9$     & Es$_7^4\;$ = 
          \,cn\,dn\,\sn\,($1-\frac{11}{5}\,\cn^4$)                   & 2\K \\
8   & 36   & L$_{10}$  & Ec$_8^4\;$ = 
          \,$1-6\,\cn^4+\frac{39}{7}\,\cn^8$                         & 2\K \\
9   & 45   & L$_{11}$  & Ec$_9^5\;$ = 
          \,cn\,($1-\frac{22}{5}\,\cn^4+\frac{11}{3}\,\cn^8$)        & 4\K \\
10  & 55   & R$_{12}$  & Es$_{10}^5$ = 
          \,dn\,\sn\,($1-\frac{26}{3}\,\cn^4+\frac{221}{21}\,\cn^8$) & 4\K \\
11  & 66   & R$_{13}$  & Es$_{11}^6$ = 
          \,cn\,dn\,\sn\,($1-6\,\cn^4+\frac{19}{3}\,\cn^8$)          & 2\K \\
12  & 78   & L$_{14}$  & Ec$_{12}^6$ =
          ($1-13\,\cn^4+\frac{221}{7}\,\cn^8
                       -\frac{221}{11}\,\cn^{12}$)                   & 2\K \\
13  & 91   & L$_{15}$  & Ec$_{13}^7$ = 
          \,cn\,($1-9\,\cn^4+19\,\cn^8
                       -\frac{437}{39}\,\cn^{12}$)                   & 4\K \\
14  & 105  & R$_{16}$  & Es$_{14}^7$ =  
          \,dn\,\sn\,($1-17\,\cn^4+51\,\cn^8
                       -\frac{425}{11}\,\cn^{12}$)                   & 4\K \\
15  & 120  & R$_{17}$  & Es$_{15}^8$ = 
          \,cn\,dn\,\sn\,($1-\frac{57}{5}\,\cn^4+\frac{437}{15}\,\cn^8
                       -\frac{1311}{65}\,\cn^{12}$)                  & 2\K \\
\hline
\end{tabular}
}{~Numbers $n$ and parameter values $\epsilon_n$ for the isochronous 
bifurcations of the A orbit in the homogeneous quartic oscillator (Q4)
potential; names O$_\sigma$ of bifurcated orbits; Lam\'e polynomials for 
their $x$ motion, and their periods $P$. The Jacobi elliptic functions sn,
cn, and dn are given in short notation as $\sn=\sn(z,k)$, etc., with 
$z=y_0t$, where $y_0$ and the modulus $k$ are given in \eq{yaq4}. (See 
text for a special comment on orbits B$_3$ and C$_4$.)
}

\noindent
orbits of type B 
and vice versa, and the orbits of type C are mapped onto themselves. This 
is seen easily in the Lam\'e polynomial describing the B orbit, 
Ec$_1^1(z)$ = cn$(z,k)$, which is proportional to the function \eq{yaq4} 
describing the A orbit. 

\Figurebb{plamorb}{-10}{276}{559}{566}{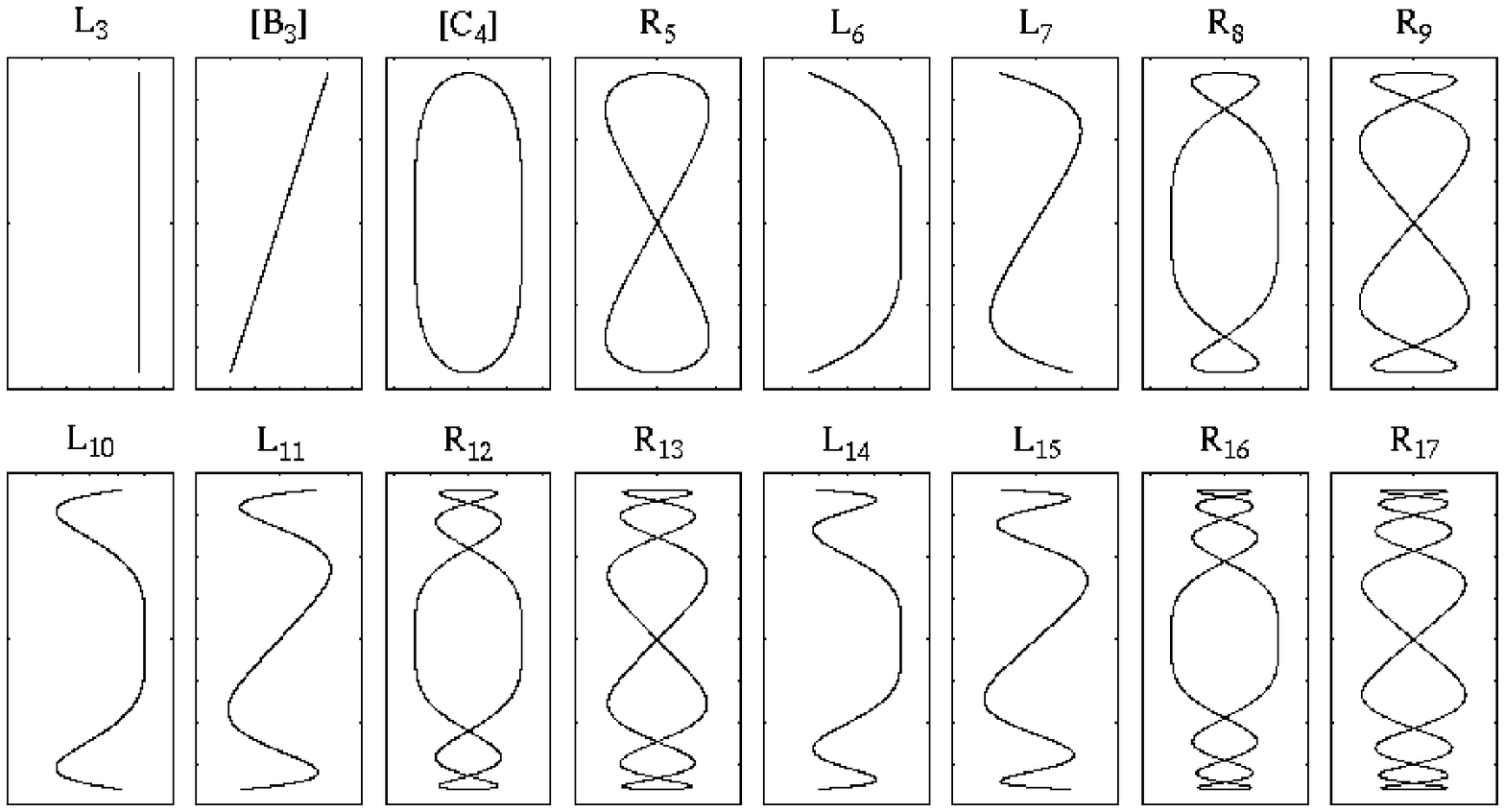}{7.8}{16.6}{
Shapes of orbits born at the isochronous bifurcations of orbit A in the 
Q4 potential. {\it Vertical direction:} $y_A(t)$ given by \eq{yaq4};
{\it horizontal direction:} $x(t)$ given by the Lam\'e polynomials 
listed in \tab{q4lame} (with scaled normalizations to fit the shapes 
into the frames).
}

The scaling properties of these orbits and their evolution away from the 
bifurcation values $\epsilon_\sigma$ are more difficult to analyze than 
in the HH and H4 potentials and will be discussed elsewhere \cite{mmmb}.

We now discuss period-doubling bifurcations of the A orbit in the Q4
potential. There is a series of trivial period doublings which just 
involve the second iterates of the orbit A and the orbits born at its 
isochronous bifurcations and listed in \tab{q4lame}. The non-trivial 
period doublings occur when $\trMA=-2$, which leads with \eq{trmaq4} 
to the critical values
\be
\epsilon = 2\,p\,(p+1)+3/8\,, \qquad p = 0,1,2,\dots 
\ee
This is exactly one of the conditions \cite{inc2} for the existence 
of period-8$\K$ solutions of \eq{lameq4}, namely that obtained by 
inserting $n=(4p+1)/2$ into Eq.\ \eq{hn}. The solutions are the 
algebraic Lam\'e functions, given (up to $p=3$) in \tab{q4lame2} below. 
These bifurcations are of the island-chain type (see, e.g., Ref.\ 
\cite{ssun}): the quantity tr\,M of the second iterate of orbit A -- 
let us call it orbit A$^2$ -- touches the value $+2$, but the orbit 
A$^2$ remains stable on either side. At the bifurcation, two 
doubly-degenerate orbits are born, one stable and one unstable. The 
situation is illustrated in \fig{q4trmt} around the bifurcation at 
$\epsilon=4+3/8$ ($p=1$). The unstable new orbit is here called 
F$_{10}$, and the stable new orbit is called P$_9$. There shapes, 
together with those born at the other period doublings listed in 
\tab{q4lame2}, are shown in \fig{plamorb2} below. Their degenerate 
symmetry partners are called F$'_\sigma$ for the librating orbits 
(obtained by reflecting the F$_\sigma$ orbits at the $x$ axis) and 
P$'_\sigma$ for the rotating orbits (obtained by time reversal of the 
P$_\sigma$ orbits).

According to the theory of Ince \cite{inc2}, the algebraic Lam\'e
functions of period 8$\K$ are one exceptional case where two 
independent periodic solutions can coexist for the same critical
value of $h$. These are the functions Ec$_{2p+1/2}^{\,p+1/2}$ and
Es$_{2p+1/2}^{\,p+1/2}$ defined in Eqs.\ \eqq{ecalg}{esalg}. As we see 
from \tab{q4lame2}, they correspond to the unstable orbits of type
F$_\sigma$ and F$'_\sigma$. In contrast to the degenerate pairs of 
bifurcated orbits in the HH and H4 potentials, which are represented by 
one and the same periodic Lam\'e function, the pairs F$_\sigma$ and 
F$'_\sigma$ are here given by two linearly independent functions. With 
this, however, the number of independent solutions of the second-order 
differential equation \eq{lameq4} is exhausted. Therefore the other pair 
of stable orbits of type P$_\sigma$ and P$'_\sigma$ born at the period 
doublings cannot be given by any new independent solutions. Indeed, we 
see from \tab{q4lame2} that the orbits P$_\sigma$ and P$'_\sigma$ are
 
\Figurebb{q4trmt}{-40}{70}{785}{525}{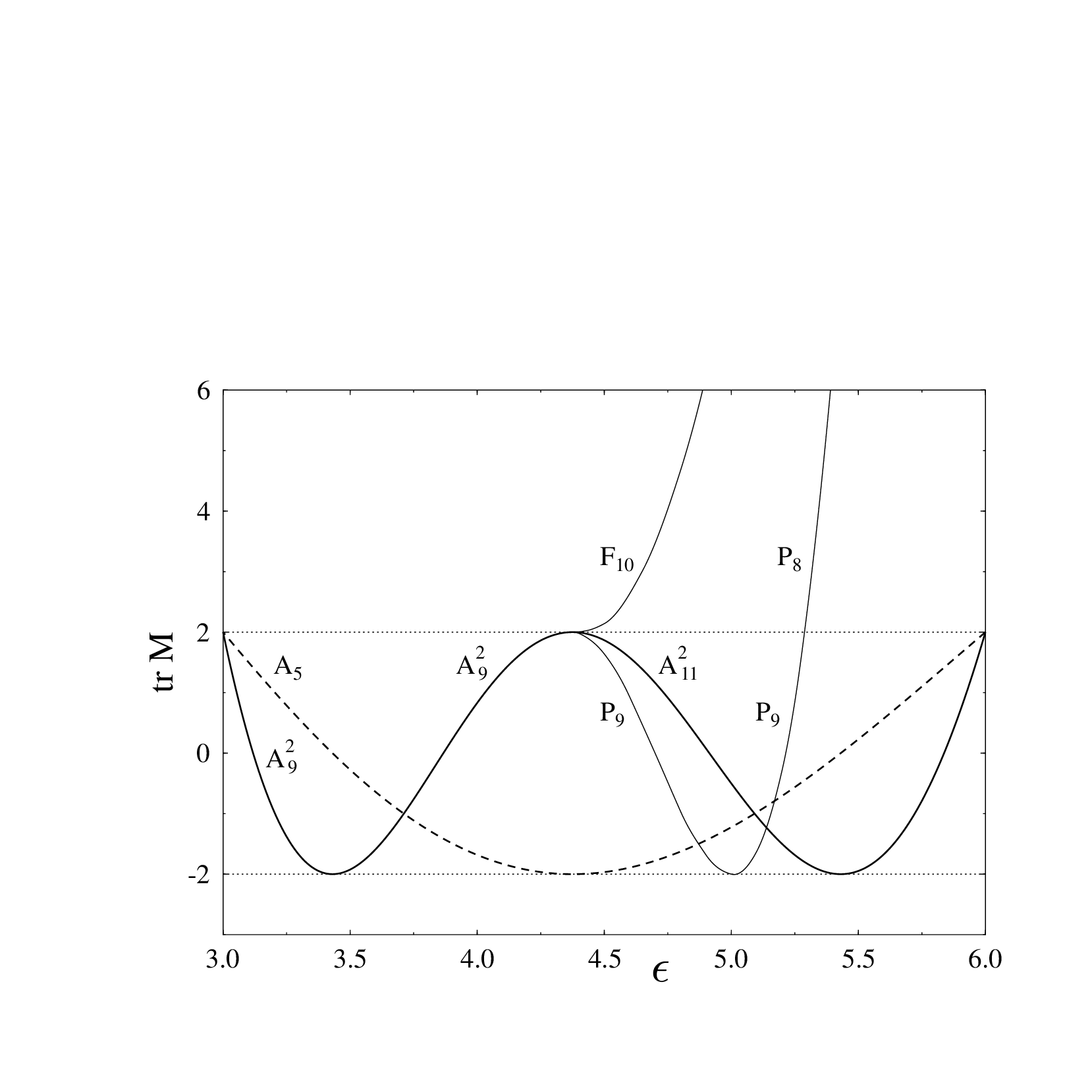}{7.5}{16.6}{
Stability discriminant $\trM$ of period-2 orbits in the Q4 potential
around a period-doubling bifurcation of island-chain type at
$\epsilon=4.375$. A$^2$ is the second iterate of the primitive orbit 
A$_5$ (shown by a dashed line); its Maslov index increases from 9 to 
11 at the bifurcation. F$_{10}$ and P$_9$ are the genuine period-2 
orbits bifurcating from it; each has a discrete degeneracy of two 
(see text and \tab{q4lame2}); P$_9$ bifurcates at $\epsilon\simeq 5.3$ 
and becomes P$_8$.
}

\Table{q4lame2}{16.6}{
\begin{tabular}{|r|r|l|l|}
\hline
$n\!$ & $\epsilon_n\!$&O$_\sigma\!$&algebraic Lam\'e function for $x(t)$\\
\hline
$\frac12\!$ & $\frac38\!$ & F$_6\!$ & Ec$_{1/2}^{1/2}$ = $
                   \!\sqrt{\dn+\cn}$ \\
      &     & F$'_6\!$ & Es$_{1/2}^{1/2}$ = $\!\sqrt{\dn-\cn}$ \\
      &     & P$_7\!$  & $\sqrt{\dn+\cn}\,+\sqrt{\dn-\cn}$ \\
      &     & P$'_7\!$ & $\sqrt{\dn+\cn}\,-\sqrt{\dn-\cn}$ \\
\hline
$\frac52\!$ & $4\frac38\!$ & F$_{10}\!$ & Ec$_{5/2}^{3/2}$ = $
              \!\sqrt{\dn+\cn}\,(1-\frac47\,\sn^2-\frac87\,\dn\,\cn)$\\
      &     & F$'_{10}\!$ & Es$_{5/2}^{3/2}$ = $\!\sqrt{\dn-\cn}\,
                                (1-\frac47\,\sn^2+\frac87\,\dn\,\cn)$\\
      &     & P$_9\!$  & Ec$_{5/2}^{3/2}\;+\;$Es$_{5/2}^{3/2}$ \\
      &     & P$'_9\!$ & Ec$_{5/2}^{3/2}\;-\;$Es$_{5/2}^{3/2}$ \\
\hline
$\frac92\!$ & $12\frac38\!$ & F$_{14}\!$ & Ec$_{9/2}^{5/2}$ = $
      \!\sqrt{\dn+\cn}\,
      (1-\frac{36}{13}\,\sn^2+\frac{16}{13}\,\sn^4-\frac{8}{13}\,\dn\,\cn)$\\
      &     & F$'_{14}\!$ & Es$_{9/2}^{5/2}$ = $\!\sqrt{\dn-\cn}\,
      (1-\frac{36}{13}\,\sn^2+\frac{16}{13}\,\sn^4+\frac{8}{13}\,\dn\,\cn)$\\
      &     & P$_{13}\!$  & Ec$_{9/2}^{5/2}\;+\;$Es$_{9/2}^{5/2}$\\
      &     & P$'_{13}\!$ & Ec$_{9/2}^{5/2}\;-\;$Es$_{9/2}^{5/2}$\\
\hline
$\frac{13}2\!$ & $24\frac38\!$ & F$_{18}\!$ & Ec$_{13/2}^{7/2}$ =$
              \,\sqrt{\dn+\cn}\,
              [1-\frac{1304}{347}\,\sn^2+\frac{1200}{347}\,\sn^4
              -\frac{320}{347}\,\sn^6+\frac{272}{347}\,\dn\,\cn\,
              (1-\frac{40}{17}\,\cn^4)]\!$\\
      &     & F$'_{18}\!$ & Es$_{13/2}^{7/2}$ =$\,\sqrt{\dn-\cn}\,
              [1-\frac{1304}{347}\,\sn^2+\frac{1200}{347}\,\sn^4
              -\frac{320}{347}\,\sn^6-\frac{272}{347}\,\dn\,\cn\,
              (1-\frac{40}{17}\,\cn^4)]\!$\\
      &     & P$_{17}\!$  & Ec$_{13/2}^{7/2}\;+\;$Es$_{13/2}^{7/2}$\\
      &     & P$'_{17}\!$ & Ec$_{13/2}^{7/2}\;-\;$Es$_{13/2}^{7/2}$\\
\hline
\end{tabular}
}{~Numbers $n$ and parameter values $\epsilon_n$ of the four lowest 
non-trivial period-doubling bifurcations of the A orbit in the homogeneous 
quartic oscillator. Given are also the names O$_\sigma$ of the four 
topologically different orbits born simultaneously at these bifurcations,
and the (linear combinations of) algebraic Lam\'e functions with period 
8$\K$ that describe their $x$ motion. (The same short notation for the 
Jacobi elliptic functions sn, dn and cn is used as in \tab{q4lame}.)
} 

\noindent
given by the two independent linear combinations 
Ec$_{2p+1/2}^{p+1/2}\,\pm$ Es$_{2p+1/2}^{p+1/2}$ which were constructed 
by Erd\'elyi \cite{erd2} to have the same symmetry properties as those 
of the 2$\K$ and 4$\K$ periodic Lam\'e functions, as discussed in Sec.\
\ref{lamsec}.

We thus have found the interesting result -- which was new to us -- that 
the stable and unstable pairs of orbits born at a period-doubling 
bifurcation of island-chain type are mutually linear combinations of each 
other. It follows from our above arguments that this result must hold for 
all Hamiltonians with the double reflection symmetry C$_{2v}$. 

\Figurebb{plamorb2}{5}{351}{560}{491}{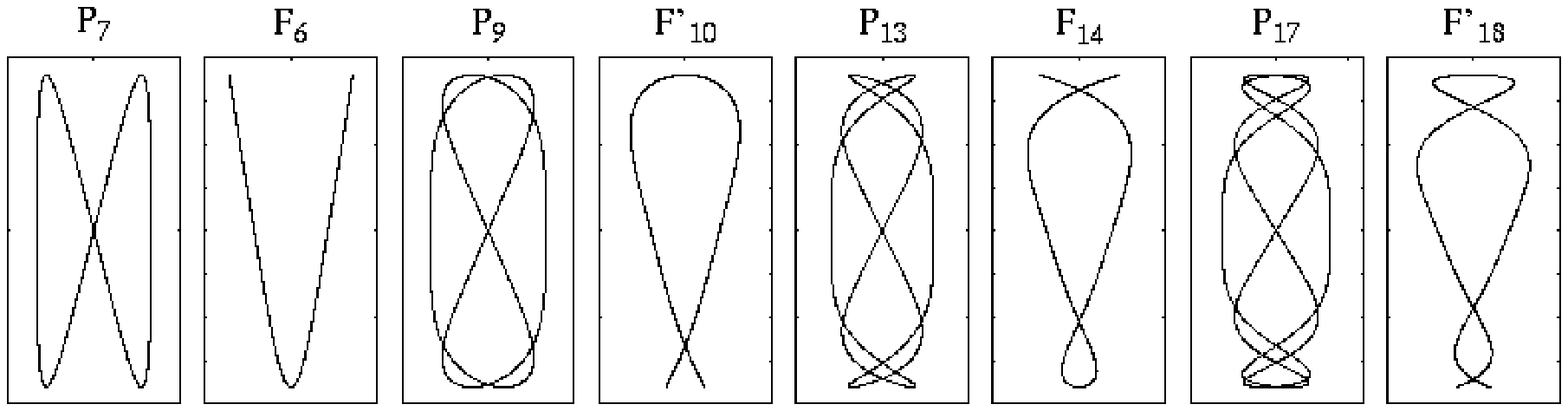}{4}{16.6}{
Same as \fig{plamorb}, but for orbits born at period-doubling 
bifurcations of orbit A in the Q4 potential. The motion $x(t)$ is given 
by the algebraic Lam\'e functions listed in \tab{q4lame2}.
}
 
\Figurebb{q4doub}{-35}{190}{559}{655}{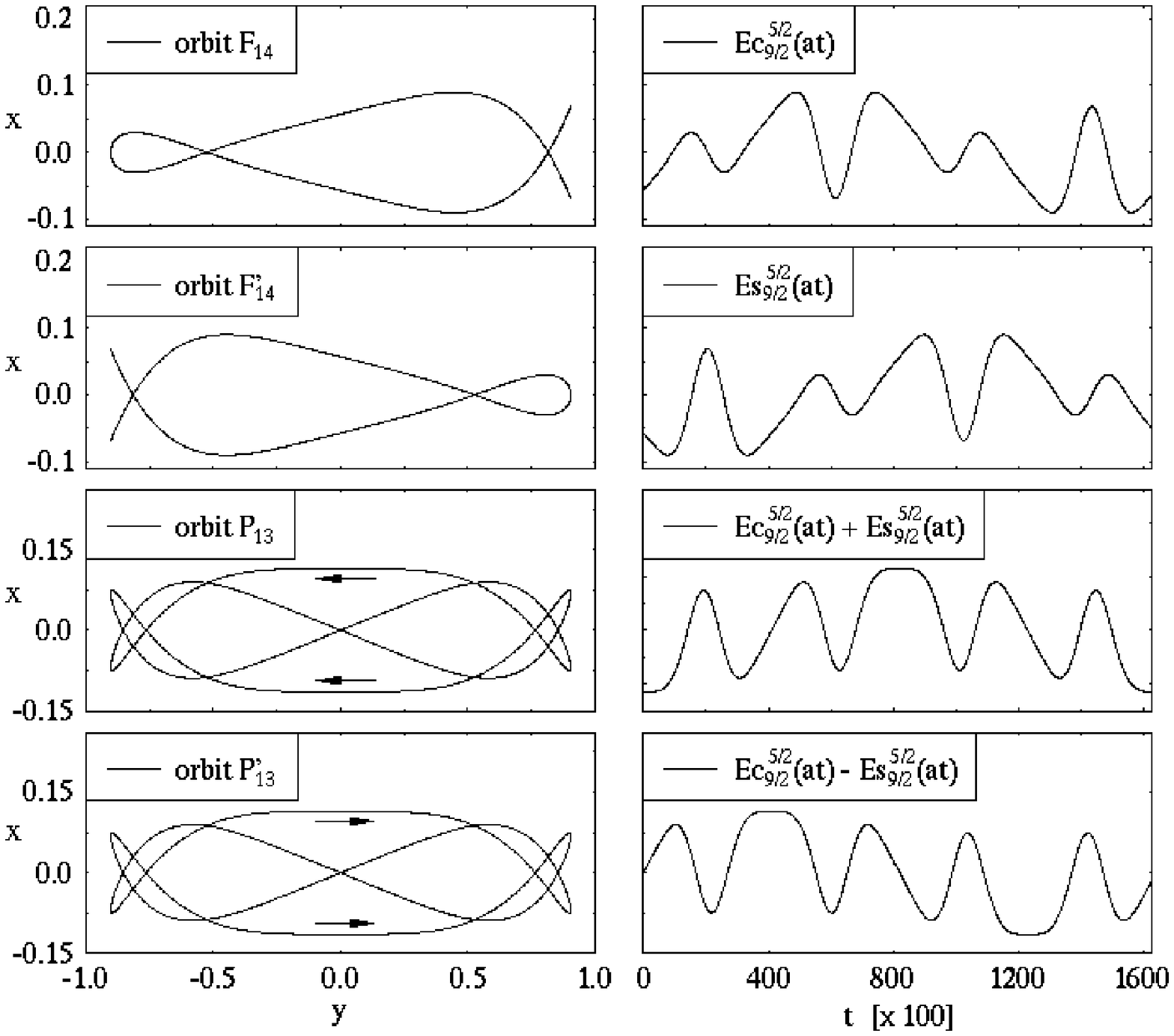}{11.5}{16.6}{
The four orbits born at the period-doubling bifurcation at 
$\epsilon=12.375$. {\it Left:} their shapes in the $(x,y)$ 
plane; {\it right:} algebraic Lam\'e functions describing their 
motion $x(t)$.
}

In \fig{q4doub} we illustrate the situation for the four orbits born at
the bifurcation at $\epsilon=12+3/8$ ($p=2$). In the left panels, their
shapes $x(y)$ are shown; in the right panels we plot the algebraic
Lam\'e functions (and their linear combinations) which describe the
motion $x(t)$ of the respective orbits. Looking merely at the shapes of 
these orbits, the pairwise linear dependence of their $x$ motion is not 
obvious at all.


\section{Summary and conclusions}
\label{susec}

\noindent
We have investigated cascades of isochronous pitchfork bifurcations of
straight-line lib\-rational orbits in two-dimensional potentials. The
linearized equation of the $x$ motion transverse to these orbits, determining
their stability, can be written in the form of the Lam\'e equation. Its
eigenvalues correspond to the bifurcation values of the system parameter,
given also by the condition tr\,M$=+2$ for the stability discriminant of
the straigh-line orbit, and 
its eigenfunctions describe the $x$ motion of the new orbits born at the 
bifurcations. These eigenfunctions are the periodic Lam\'e functions of 
period $2\K$ or $4\K$, where $\K$ is the complete elliptic integral
determining the period of the parent orbit at the corresponding 
bifurcation. In potentials with C$_{2v}$ symmetry, the solutions occur 
alternatingly as Lam\'e functions of period 2{\bf K} and 4{\bf K}, 
respectively. When this symmetry is absent, the $4\K$ periodic solutions
describe the orbits born at period-doubling pitchfork bifurcations.

We have shown numerically that the periodic Lam\'e functions describe very 
accurately the shapes of the bifurcated orbits obtained from a numerical 
integration of the equations of motion, as long as the amplitude of their 
$x$ motion remains small, i.e., as long as one is not too far from the 
bifurcation point. Exploiting the energy conservation in the 
H\'enon-Heiles type potentials HH and H4 and the known symmetries of the 
Lam\'e functions, we can predict the propagation of the new orbits up to 
the critical saddle-point energy where they have all become unstable and 
the system is highly chaotic. We thus have found an analytical description 
of an infinite series of unstable periodic orbits in chaotic systems. 

In the homogeneous quartic oscillator (Q4) potential, the series expansions
of the periodic Lam\'e functions terminate and they become finite 
polynomials. In this potential we have also analyzed solutions of period 
$8\K$ which occur at period-doubling bifurcations of the straight-line 
orbits of island-chain type. The two pairs of orbits born thereby are 
represented by two independent sets of orthogonal periodic solutions of the
Lam\'e equation, which here are identified with the so-called algebraic 
Lam\'e functions that can again be given in a closed form.

Similar cascades of pitchfork bifurcations have also been discussed in
connection with the diamagnetic Kepler problem represented by hydrogen 
atoms in strong magnetic fields \cite{maod,main,frwi}. Expressing the
Hamiltonian in (scaled) semiparabolic coordinates $(u,v)$, the effective 
potential for orbits with angular momentum $L_z=0$ (where the $z$ axis
is the direction of the external magnetic field) becomes similar to the 
H4 and Q4 potentials discussed here (although it contains only quadratic 
and sixth-order terms in the coordinates). Since physically the $(u,v)$ 
coordinates are positive definite, the periodic orbits in the diamagnetic
Kepler problem correspond to the half-orbits of the H4 and the Q4 
potentials. There exist straight-line librating orbits, corresponding 
to oscillations of the electron along the symmetry axis, which bifurcate 
infinitely many times as the energy of the electron approaches the 
ionization threshold. The stability of these linear orbits is given by the 
Mathieu equation, which is analogous to the Lam\'e equation \eq{lameq} but 
with the function sn$^2(z,k)$ replaced by $\cos(2z)$. Its periodic 
solutions are the periodic Mathieu functions se$_m$ and ce$_m$ which have 
properties completely analogous to those of the periodic Lam\'e functions, 
and were actually studied in detail by Ince \cite{inc3} prior to his 
investigations of the Lam\'e functions. The topology of the Mathieu 
functions and of the bifurcated orbits described by them is exactly the 
same as for the Q4 and H4 potentials described here. In particular, the 
so-called ``balloon'' orbits B$_n$ and ``snake'' orbits S$_n$ with $n=1,$ 
2, $\dots$ \cite{maod} correspond exactly to the alternating sequence of 
halves of the orbits R$_5$, R$_9$, $\dots$ and L$_7$, L$_{11}$, $\dots$ shown 
in \fig{plamorb}. (The other orbits, born unstable at the bifurcations, 
were not considered in \cite{maod} since they do not pass through the 
centre.) We believe that our analysis, applied in terms of the Mathieu 
functions, may be useful for further investigations of the bifurcations 
occurring in the diamagnetic Kepler problem. 

The knowledge of the analytical properties of the bifurcated orbits will 
be useful in the application of the periodic orbit theory to the potentials 
studied here. First steps in this direction have been quite successful 
\cite{pert,hhun,hh1}, but the orbits bifurcated from the A orbit were not 
considered. Their incorporation into the semiclassical trace formula is
the object of further work in progress. We expect the bifurcated orbits, 
in particular, to play an important role in the semiclassical analysis of 
resonances above the barriers in H\'enon-Heiles type or similar potentials.

\bs
\bs

{\bf Acknowledgments}

\ms

\noindent
We are grateful to S Fedotkin and A Magner for very stimulating 
discussions and critical comments. Valuable comments by M Sieber and
H Then are highly appreciated. We also acknowledge financial support 
by the Deutsche Forschungsgemeinschaft.

\end{document}